\documentclass[10pt, a4paper]{article}
\pdfoutput=1
\usepackage[utf8]{inputenc}
\usepackage[T1]{fontenc}
\usepackage{lmodern, textcomp}
\usepackage[british]{babel}
\usepackage{csquotes}

\usepackage{eurosym, xspace}
\usepackage[nottoc]{tocbibind}
\usepackage{graphicx, subcaption, float}
\usepackage{marginnote}
\usepackage{xparse}
\usepackage{authblk}
\usepackage{changepage}

\usepackage[usenames, dvipsnames, svgnames, table]{xcolor}

\usepackage[backend=bibtex8, citestyle=numeric-comp, maxbibnames=99,
			sorting=none, sortcase=false, sortcites=true, giveninits=true]{biblatex}
\input{arxiv.def}

\usepackage{amsmath, amsfonts, amssymb}
\usepackage{cancel}
\usepackage{bm, bbm}

\usepackage{maths_min}

\usepackage{hyperxmp}
\usepackage[pdftex, breaklinks=true, linktocpage,
	colorlinks=true, urlcolor=blue, linkcolor=blue, citecolor=red,
	pdfauthor={Harold Erbin}
]{hyperref}
\usepackage[all]{hypcap}

\DeclareUnicodeCharacter{00A0}{~}
\DeclareUnicodeCharacter{202F}{~}

\newcommand{\email}[1]{\href{mailto:#1}{\nolinkurl{#1}}}
\newcommand{\emailfoot}[1]{\thanks{\email{#1}}}

\newcounter{draftcommentcnt}
\NewDocumentCommand{\draftcomment}{s O{red} m}{%
	\def\margnote{\IfBooleanTF{#1}{\marginnote}{\marginpar}}%
	\stepcounter{draftcommentcnt}%
	\textcolor{#2}{#3}%
	\margnote{\textcolor{#2}{$\Leftarrow$ \arabic{draftcommentcnt}}}%
}

\numberwithin{equation}{section}

\usepackage[sort&compress, english]{cleveref}

\usepackage[heightrounded,
	top=4cm, bottom=4cm, left=3cm, right=3cm,
]{geometry}

\input{arxiv.def}

\hypersetup{
	pdftitle={Conformality of 1/N corrections in SYK-like models},
}

\title{Conformality of $1/N$ corrections in SYK-like models}

\author[1]{Stéphane Dartois\emailfoot{stephane.dartois@outlook.com}}
\affil[1]{Laboratoire de Physique Théorique, CNRS UMR 8627, Université Paris XI, 91405 Orsay Cedex, France, \textsc{Eu}.}

\author[2]{Harold Erbin\emailfoot{erbin@lpt.ens.fr}}
\affil[2]{\textsc{Lpt}, Département de physique de l'\textsc{Ens}, École normale supérieure, \textsc{Upmc} Univ. Paris 06, \textsc{Cnrs}, \textsc{Psl} Research University, 75005 Paris, France}
\affil[2]{Sorbonne Universités, \textsc{Upmc} Univ. Paris 06, École normale supérieure, \textsc{Cnrs}, \textsc{Lpt}, 75005 Paris, France}

\author[3]{Swapnamay Mondal\emailfoot{swapno@lpthe.jussieu.fr}}
\affil[3]{Sorbonne Universités, \textsc{Upmc} Univ Paris 06, \textsc{Umr} 7589, \textsc{Lpthe}, F-75005, Paris, France}
\affil[3]{\textsc{Cnrs}, \textsc{Umr} 7589, \textsc{Lpthe}, F-75005, Paris, France}

\begin{document}

\maketitle

\begin{abstract}
The Sachdev--Ye--Kitaev is a quantum mechanical model of $N$ Majorana fermions which displays a number of appealing features -- solvability in the strong coupling regime, near-conformal invariance and maximal chaos -- which make it a suitable model for black holes in the context of the AdS/CFT holography.
In this paper, we show for the colored SYK model and several of its tensor model cousins that the next-to-leading order in the large $N$ expansion preserves the conformal invariance of the $2$-point function in the strong coupling regime, up to the contribution of the pseudo-Goldstone bosons due to the explicit breaking of the symmetry and which are already seen in the leading order $4$-point function.
We also comment on the composite field approach for computing correlation functions in colored tensor models.
\end{abstract}

\newpage

\hrule
\pdfbookmark[1]{\contentsname}{toc}
\tableofcontents
\bigskip
\hrule

\section{Introduction}
\label{sec:introduction}

In a series of seminal conferences~\cite{Kitaev:2015:HiddenCorrelationsHawking, Kitaev:2015:SimpleModelQuantum-1, Kitaev:2015:SimpleModelQuantum-2} Kitaev brought attention to the -- now so called -- Sachdev--Ye--Kitaev (SYK) model which displays a set of appealing features in the context of holography, for which a detailed account has been given in~\cite{Maldacena:2016:CommentsSachdevYeKitaevModel}.
This model -- a simplification of a previous one by Sachdev and Ye~\cite{Sachdev:1993:GaplessSpinFluidGround} -- corresponds to a quantum mechanical system of $N$ Majorana fermions (possibly organized in different families~\cite{Gross:2017:GeneralizationSachdevYeKitaev}) with an interaction of order $q$ with Gaussian random couplings.

The first key property is that it is solvable at large coupling (or equivalently large time or infrared regime) in the large $N$ limit.
This is very precious since systems that are tractable in the large coupling regime are very scarce.
Moreover, in this infrared limit, the system displays an approximate conformal symmetry.
Conformal invariance in one dimension is equivalent to reparametrization invariance, and thus is infinite-dimensional which leads to many simplifications: in particular the system at zero and finite temperature are easily related in this regime.
This symmetry is spontaneously broken and leads to Goldstone bosons.
Since the full action breaks the symmetry explicitly but slightly, these are in fact pseudo-Goldstone, their dynamics being described by the Schwarzian action.
The latter are responsible for the last property of the model: the Lyapunov exponent, which measures the chaos in the system, reaches the maximal bound proposed in~\cite{Maldacena:2016:BoundChaos} and thus the system is maximally chaotic.
All together these properties point towards a (near) $\group{AdS}_2 / \group{CFT}_1$ interpretation of the model (see~\cite{Almheiri:2015:ModelsAdS2Backreaction, Jensen:2016:ChaosAdS2Holography, Engelsoy:2016:InvestigationAdS2Backreaction, Maldacena:2016:ConformalSymmetryIts} for references on near $\group{AdS}_2$).
In gravitational theories the maximally chaotic objects are black holes: hence one can expect that the bulk geometry dual of the SYK model corresponds to near-horizon geometry of black holes.
The fact that one can access the strong coupling regime offers an inestimable window on the quantum properties of black holes.

Another interesting property is its equivalence with random tensor fields theories in the large $N$ limit, as was pointed in~\cite{Witten:2016:SYKLikeModelDisorder} (some selected references on tensor models include~\cite{Gurau:2011:1NExpansionColored, Gurau:2011:1NExpansionColoredAnyD, Bonzom:2011:CriticalBehaviorColored, Gurau:2011:GeneralizationVirasoroAlgebra, Gurau:2012:Complete1NExpansion, Bonzom:2012:RandomTensorModels, Gurau:2012:ColoredTensorModels, Gurau:2012:SchwingerDysonEquations, Bonzom:2013:RevisitingRandomTensor, Rivasseau:2014:TensorTrackIII, Gurau:2016:RandomTensors, Rivasseau:2016:RandomTensorsQuantum}).
The Gurau--Witten model~\cite{Witten:2016:SYKLikeModelDisorder} is the simplest colored tensor model and consists in a set of $q = D + 1$ real fermionic tensor fields with $D$ indices of size $N$ transforming in the fundamental of $\group{O}(N)^{\otimes D}$, the invariance group being $\group{O}(N)^{D(D+1)/2}$ (up to a discrete factor).
Two other models of interest are the case of $q$ complex fermionic tensor fields with a $\group{U}(N)$ invariance and the so-called multi-orientable model which is given by a complex fermionic field with $D = 3$ indices~\cite{Klebanov:2017:UncoloredRandomTensors} (there is no $q$ because one is considering an uncolored tensor model~\cite{Bonzom:2012:RandomTensorModels}).
The bosonic $0$-dimensional versions of these models have been studied in~\cite{Carrozza:2016:ONRandomTensor, Gurau:2011:1NExpansionColored, Gurau:2011:1NExpansionColoredAnyD, Dartois:2014:1NExpansionMultiorientable, Raasakka:2015:NexttoleadingOrderLarge, Tanasa:2016:MultiOrientableRandomTensor}.
The main simplification in these models occur because the randomness is moved in the fields and there is a single (fixed) coupling constant.
While it is necessary to average over the random couplings by performing the Gaussian integration over them (quenching), implying that one describes a thermodynamical ensemble, the tensor models feature a unique fixed coupling constant and represent a genuine quantum system~\cite{Witten:2016:SYKLikeModelDisorder}.
Moreover the combinatorics and renormalization properties have been largely studied and one can make use of all the tools already developed.

The disordered and tensor SYK models have been extended in several directions:
higher dimensions and lattices~\cite{Gu:2016:LocalCriticalityDiffusion, Berkooz:2017:HigherDimensionalGeneralizations, Turiaci:2017:2dQFTAnalog, Jian:2017:SolvableSYKModels, Berkooz:2017:CommentsRandomThirring, Jian:2017:ModelContinuousThermal, Peng:2017:VectorModelsGeneralized, Narayan:2017:SYKlikeTensorModels, Khveshchenko:2017:ThickeningSickeningSYK},
$N = 1, 2$ supersymmetry~\cite{Fu:2017:SupersymmetricSYKModels, Peng:2016:SupersymmetricSYKlikeTensor} (see~\cite{Anninos:2016:DisorderedQuiversCold} for a related system with $N = 4$),
non-quenched disorder~\cite{Michel:2016:FourpointFunctionIOP, Nishinaka:2016:NoteSachdevYeKitaevModel, Gurau:2017:QuenchedEqualsAnnealed}.
Various properties have been studied in the last year:
spectrum and thermodynamical properties~\cite{Polchinski:2016:SpectrumSachdevYeKitaevModel, Cotler:2017:BlackHolesRandom, Fu:2016:NumericalStudyFermion, Liu:2016:DisorderSachdevYeeKitaevModel, Banerjee:2017:SolvableModelDynamical, Gurau:2017:Complete1NExpansion, GarciaGarcia:2017:AnalyticalSpectralDensity, Bi:2017:InstabilityNonFermiLiquid, Chen:2017:TunableQuantumChaos, Chen:2017:CompetitionChaoticNonChaotic, Song:2017:StronglyCorrelatedMetal},
correlation functions~\cite{Polchinski:2016:SpectrumSachdevYeKitaevModel, Jevicki:2016:BiLocalHolographySYK, Jevicki:2016:BiLocalHolographySYK-1, Bonzom:2017:DiagrammaticsColoredSYK, Gross:2017:BulkDualSYK, Bagrets:2017:PowerlawOutTime, Gurau:2017:imathEpsilonPrescription},
dynamics of the Goldstone bosons~\cite{Maldacena:2016:CommentsSachdevYeKitaevModel, Bagrets:2016:SachdevYeKitaevModelLiouville, Bagrets:2017:PowerlawOutTime, Stanford:2017:FermionicLocalizationSchwarzian},
relation with matrix models (for both the disordered and tensor versions)~\cite{Cotler:2017:BlackHolesRandom, You:2017:SachdevYeKitaevModelThermalization, Krishnan:2017:QuantumChaosHolographic, GarciaGarcia:2017:AnalyticalSpectralDensity, Li:2017:SupersymmetricSYKModel, Ferrari:2017:LargeLimitPlanar, Krishnan:2017:RandomMatricesHolographic},
transport properties~\cite{Gu:2016:LocalCriticalityDiffusion, Davison:2016:ThermoelectricTransportDisordered, Gu:2017:EnergyDiffusionButterfly}, renormalization and phases~\cite{Bi:2017:InstabilityNonFermiLiquid}.
Experimental realizations have been proposed in~\cite{GarciaAlvarez:2016:DigitalQuantumSimulation, Danshita:2016:CreatingProbingSachdevYeKitaev, Pikulin:2017:BlackHoleChip, Chew:2017:ApproximatingSachdevYeKitaevModel}.

It was shown in~\cite{Polchinski:2016:SpectrumSachdevYeKitaevModel, Maldacena:2016:CommentsSachdevYeKitaevModel} that the next-to-leading order (NLO) correction in the coupling constant breaks explicitly the conformal invariance in the leading order (LO) in the large $N$ expansion.
The problem we address in this paper is the reversed one, i.e.\ is the conformal symmetry explicitly broken in the NLO in $N$ for the LO in the coupling constant?
We consider this question in the models mentioned above: the colored\footnotemark{} SYK model with disorder, and the real, complex and multi-orientable SYK tensor models.
\footnotetext{%
	This study is restricted to the colored SYK model (already discussed in~\cite{Gross:2017:GeneralizationSachdevYeKitaev, Bonzom:2017:DiagrammaticsColoredSYK, Gurau:2017:QuenchedEqualsAnnealed}) because the combinatorics of graphs involving (anti)symmetric tensor is notoriously difficult and was one of the reason for the lack of progress in tensor models, until Gurau solved this problem by introducing colors~\cite{Gurau:2011:1NExpansionColored}.
}%
We find that in the first three models the NLO $2$-point function is compatible with conformal symmetry and thus should scale in the same way as the LO $2$-point function.
This means that in the infrared the dimension of the fermions is not modified by the first subleading correction in the large $N$ expansion.
This finding may have some implications for the construction of the bulk dual of SYK which has started in~\cite{Gross:2017:BulkDualSYK} (see also~\cite{Maldacena:2016:CommentsSachdevYeKitaevModel, Jevicki:2016:BiLocalHolographySYK, Jensen:2016:ChaosAdS2Holography} and~\cite{Mandal:2017:CoadjointOrbitAction, Das:2017:ThreeDimensionalView} for other proposals).
Our method consists in analysing the transformation properties of the NLO $2$-point function from the Schwinger--Dyson equation (the Feynman graphs contributing at this order have been studied in~\cite{Bonzom:2017:DiagrammaticsColoredSYK}, see also~\cite{Raasakka:2015:NexttoleadingOrderLarge}): this is sufficient to reach our conclusions except for the multi-orientable tensor model.
In the latter case the conclusion depends on the explicit form of the NLO $2$-point function and the full analysis is outside the scope of this paper.
It is important to note that in all this paper the divergent contribution to the LO $4$-point function due to the spontaneous breaking of the conformal symmetry is implicitly excluded (as is implied by any statement in previous works about the conformality of some object)~\cite{Maldacena:2016:CommentsSachdevYeKitaevModel}: this contribution can be taken into account only by looking at the NLO in the coupling which regularizes the divergence.

A fruitful approach for computing the correlation functions and determining the structure of the graph appearing at some order in $N$ is to write the action in terms of composite fields -- to be identified with the $2$-point function and self-energy -- instead of the fundamental fermions~\cite{Jevicki:2016:BiLocalHolographySYK, Jevicki:2016:BiLocalHolographySYK-1, Maldacena:2016:CommentsSachdevYeKitaevModel, Bagrets:2016:SachdevYeKitaevModelLiouville}.
We briefly discuss in \cref{sec:fluctuations} how such an approach can be undertaken for the colored tensor models.

The structure of the paper as follows.
In \cref{sec:syk-disorder,sec:syk-colored-tensor,sec:syk-multi-tensor} we study successively the SYK model, the (real and complex) colored tensor models and the multi-orientable tensor models.
The results are discussed in \cref{sec:discussion}.
\Cref{sec:fluctuations} describes how to perform a composite field analysis for the real colored tensor model.

\section{SYK model with disorder}
\label{sec:syk-disorder}

\subsection{The model}

In this section, we consider a specific case of the colored SYK model introduced in \cite{Gross:2017:GeneralizationSachdevYeKitaev}.
This has the main advantage of simplifying the study of the combinatorics (which has been done in \cite{Bonzom:2017:DiagrammaticsColoredSYK} in details) and makes it easier to compare with the Gurau--Witten colored tensor model later described.
However, the model we study here keep all the interesting features of the usual SYK model at leading order.

\medskip

The colored SYK model we consider is a model of $q N$ real massless fermions $\psi^c_i$ where $c \in \{1\ldots q\}$, $i \in \{1\ldots N\}$, with $q$ being the color index.
This model is defined through the following Euclidean space partition function
\begin{equation}
	Z^{\text{SYK}}_{N, \lambda}
		= \int \dd\lambda \,
			\exp\left(
				- \frac{N^{q-1}}{2 \lambda^2} \sum_{\{i_k\}_{k = 1}^q}^N \lambda_{i_1\ldots i_q} \lambda_{i_1\ldots i_q}
				\right)
			\int \prod_{c = 1}^q \mathcal{D} \psi^c\,
			\e^{- \int \dd t L[\psi, \lambda]}
\end{equation}
where
\begin{equation}
	L[\psi, \lambda]
		= \frac{1}{2} \sum_{c = 1}^q \sum_{i_c = 1}^N \psi^c_{i_c}\partial_t\psi^c_{i_c}
			+ \frac{\I^{q/2}}{q!} \sum_{\{i_k\}_{k = 1}^q}^N \lambda_{i_1\ldots i_q} \prod_{c = 1}^q \psi^c_{i_c}.
\end{equation} 
No particular assumption is made on the symmetry of the random couplings $\lambda_{i_1\ldots i_q}$ and it is convenient to define
\begin{equation}
	g = \lambda^2.
\end{equation} 
The reason is that there is no need for antisymmetry on the indices here since no color appears twice in the interaction term.
Moreover, this simplifies further the combinatorics as this prohibits melonic graphs from contributing to subleading amplitudes in $1/N$ as well.


The free scalar two-point function $G_f(t_1, t_2)$ is defined, after an arbitrary choice of color $c_0$ (which is kept implicit in the notation), by 
\begin{equation}
	G_f(t_1, t_2)
		= \frac{1}{N} \Big\langle \sum_{i} T \psi^{c_0}_{i}(t_1) \psi^{c_0}_{i}(t_2) \Big\rangle_0
		= \frac{1}{2}\, \sign(t_1 - t_2),
\end{equation}
whose Fourier transform writes
\begin{equation}
	G_f(\omega) = - \frac{1}{i\omega}.
\end{equation}
We also have that,
\begin{equation}
	\Big\langle T \psi^c_{i_c}(t_1) \psi^{c'}_{i_{c'}}(t_2) \Big\rangle_0
		= \delta_{cc'} \delta_{i_c i_{c'}}\, G_f(t_1, t_2).
\end{equation}
The exact disorder averaged two-point function $G_e(t_1, t_2)$ is defined by the following relations
\begin{align}
	G_e(t_1, t_2)
		&= \frac{1}{N} \Big\langle \sum_{i} T \psi^{c_0}_{i}(t_1) \psi^{c_0}_{i}(t_2) \Big\rangle,
	\\
	\langle T \psi^c_{i_c}(t_1) \psi^{c'}_{i_{c'}}(t_2)\rangle
		&= \delta_{cc'}\delta_{i_ci_{c'}}\, G_e(t_1, t_2).
\end{align} 
The Feynman graphs of this model are made up of the following building blocks:
\begin{itemize}
	\item The vertices are $q + 1$ valent.

	\item The edges are of two types.
	The fermionic edges carry a \emph{color} label $c \in \{1\ldots q\}$.
	The disorder edges carry a $0$ label.
	Edges are labelled in such a way that no two adjacent edges have the same label (color or disorder).
	
	\item The faces are cycles made alternatively of edges labelled $0$ and $c$, for some color label.
\end{itemize}
The free energy of the colored SYK has a $1/N$ expansion of the form
\begin{equation}
	\label{eq:syk-F}
	F^{\text{SYK}}_{N, \lambda}
		= \log Z^{\text{SYK}}_{N, \lambda}
		= \sum_{\ell_m \ge 0} N^{1-\ell_m} F_{[\ell_m]}(\lambda),
\end{equation}
where $\ell_m(G)$ is a characteristic number of the Feynman graph $G$: more precisely, it is the number of multi-colored cycles of the graph $G_{\backslash 0}$ that is obtained from $G$ by contracting all edges labelled $0$.
From these considerations, we get that the exact two-point function also admits a $1/N$ expansion
\begin{equation}
	\label{eq:syk-Ge}
	G_e(t_1, t_2)
		= \sum_{\ell_m\ge 0} N^{-\ell_m} G_{[\ell_m]}(t_1, t_2).
\end{equation}

\subsection{Leading Order}

The leading order of the SYK model has been described in several works \cite{Maldacena:2016:CommentsSachdevYeKitaevModel, Polchinski:2016:SpectrumSachdevYeKitaevModel}, and the colored SYK model has been described in \cite{Gross:2017:GeneralizationSachdevYeKitaev}, therefore, we only give a very brief account and the reader may refer to the excellent presentations mentioned above for more details.
The leading order two-point function ($\ell_m = 0$), $G_{[0]}$ satisfies the following equation:
\begin{equation}
	G_{[0]}(t_1, t_2)
		= G_f(t_1, t_2) + g \int \dd t \dd t'\,
			G_f(t_1, t) \Sigma_{[0]}(t, t') G_{[0]}(t', t_2),
\end{equation}
where $\Sigma_{[0]}$ is the leading order self-energy.
The above relation is easily obtained from the usual relation between the two-point function and the self-energy
\begin{equation}
	\label{eq:SE-to-2pt-exact}
	G_e(t_1, t_2)
		= \big(G_f(t_1, t_2)^{-1} - \Sigma(t_1, t_2) \big)^{-1},
\end{equation}
where the inverse here means that the two variables functions $G_f(t_1, t_2),\ \Sigma(t_1, t_2)$ are seen as matrices for the convolution product.
The graphs appearing at leading order are the \emph{melonic} graphs, also called melon graphs (see \cite{Witten:2016:SYKLikeModelDisorder, Bonzom:2011:CriticalBehaviorColored} for a description of these graphs).
This implies that 
\begin{equation}
	\Sigma_{[0]}(t, t') = G_{[0]}(t, t')^{q-1}.
\end{equation} 
Therefore we have
\begin{equation}
	G_{[0]}(t_1, t_2)
		= G_f(t_1, t_2) + g \int \dd t \dd t'\,
			G_f(t_1, t) G_{[0]}(t, t')^{q-1} G_{[0]}(t', t_2).
\end{equation}
In Fourier space, this equation rewrites
\begin{equation}
	- \I \omega \, G_{[0]}(\omega)
		= 1 + g \, \Sigma_{[0]}(\omega) G_{[0]}(\omega).
\end{equation}

Consequently, in the infrared limit\footnotemark{} $G_{[0]} \rightarrow \bar{G}_{[0]}$, the left hand side drops and one has
\footnotetext{%
	One recovers the same results if one considers the large coupling $g$ limit.
}%
\begin{equation}
	0 = 1 + g \, \bar{\Sigma}_{[0]}(\omega)\bar{G}_{[0]}(\omega),
\end{equation} 
where $\bar{G}_{[0]}$ stands for the infrared limit of $G_{[0]}$ and $\bar{\Sigma}_{[0]}$ stands for the infrared limit of $\Sigma_{[0]}$.
In the rest of the paper, any barred quantity denotes the infrared or large coupling limit of the corresponding unbarred quantity.
In position space, we have
\begin{equation}
	g \int \dd t \, \bar{G}_{[0]}(t_1, t)^{q-1} \bar{G}_{[0]}(t, t_2)
		= - \delta(t_1 - t_2).
\end{equation}
In this regime, the $2$-point function transforms as
\begin{equation}
	\label{eq:syk-reparam-2pt}
	\bar{G}_{[0]}(\sigma, \sigma')
		= \frac{1}{|f'(t) f'(t')|^{1/q}} \, \bar{G}_{[0]}(t, t')
\end{equation} 
under reparametrizations $\sigma = f(t)$ and $\sigma' = f('t)$.

\subsection{The Next-to-Leading Order}

In this subsection, we want to investigate the Next-to-Leading Order (NLO) of the colored SYK model.

We want to study the possible corrections to the scaling dimension of the two-point function in the conformal sector.
The NLO is given by the graphs with $\ell_m = 1$ in \eqref{eq:syk-F}, which means that their contracted graphs have one multicolored cycle.
To obtain the $2$-point function $G_{\text{NLO}}$ we will first get need to describe to self-energy $\Sigma_{\text{NLO}}$ at NLO, where we defined:
\begin{equation}
	G_{\text{NLO}} := G_{[1]},
	\qquad
	\Sigma_{\text{NLO}} := \Sigma_{[1]},
\end{equation} 
see \eqref{eq:syk-Ge}.
We use the results from \cite{Bonzom:2017:DiagrammaticsColoredSYK} which classified the NLO graphs.
From this work, it is possible to write the self-energy at NLO as:
\begin{adjustwidth}{-2cm}{-2cm}
\begin{multline}
\label{eq:SE-NLO-SYK}
	\Sigma_{\text{NLO}}
		= \sum_{c \neq c_0}
			\raisebox{-6.0ex}{\includegraphics[scale = 0.8]{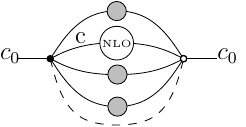}} + \sum_{l \ge 0} \sum_{\substack{c_1, c_2 \neq c_0\\ l = 0\Rightarrow c_1 = c_2}}\raisebox{-12.0ex}{\includegraphics[scale = 0.55]{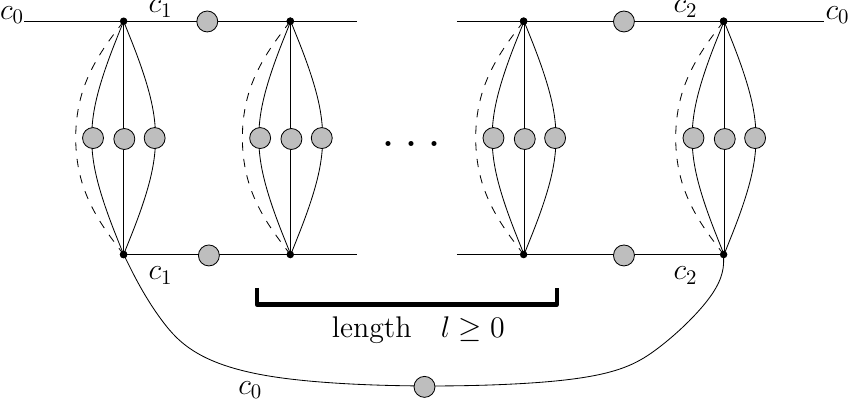}}\\
		+ \sum_{l \ge 0} \sum_{\substack{c_1, c_2 \neq c_0\\ l = 0 \Rightarrow c_1 \neq c_2}}
			\raisebox{-12.0ex}{\includegraphics[scale = 0.55]{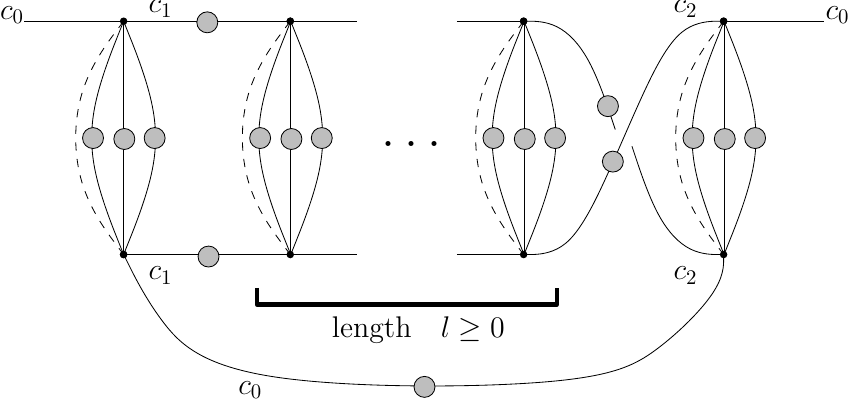}} \\
		+ \sum_{l \ge 0} \sum_{\substack{c_1, c_2, c_3 \neq c_0\\ c_1 \neq c_3 \neq c_2 \\ l = 0 \Rightarrow c_1 = c_2, c_1 \neq c_3}}
			\raisebox{-15.5ex}{\includegraphics[scale = 0.55]{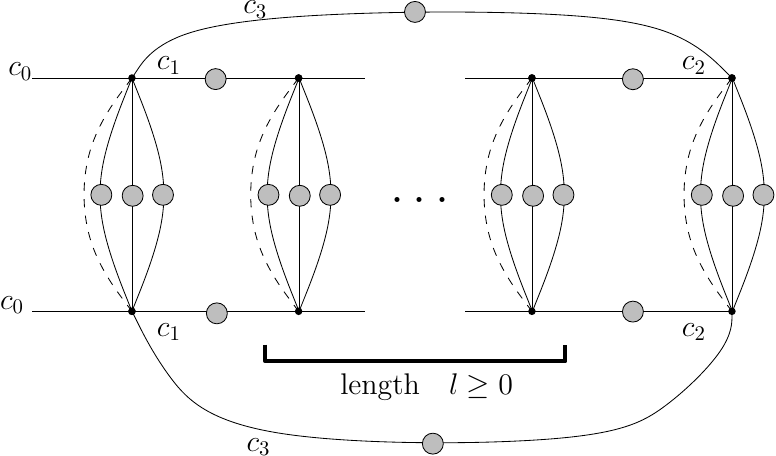}}\\
		+ \sum_{l \ge 0} \sum_{\substack{c_1, c_2, c_3 \neq c_0\\ c_1 \neq c_3 \neq c_2\\ l = 0 \Rightarrow c_1 = c_2, c_1 \neq c_3}}
			\raisebox{-15.5ex}{\includegraphics[scale = 0.55]{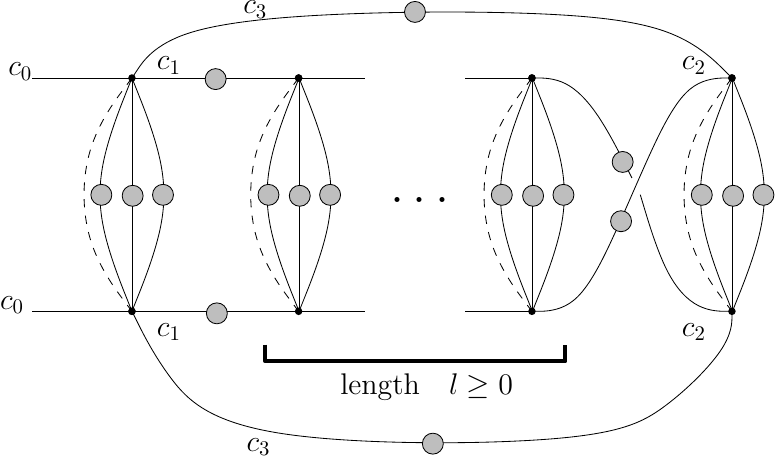}}
\end{multline}
\end{adjustwidth}
The edges with gray discs insertions represent dressed leading order propagators.

\medskip

We give a few indications on the correspondence between these terms and the graphs described in \cite{Bonzom:2017:DiagrammaticsColoredSYK}.
In the language of \cite{Bonzom:2017:DiagrammaticsColoredSYK}, the two-point function is obtained by cutting an edge of an NLO vacuum graph.
NLO vacuum graphs are ladder diagrams closed on themselves.
They can be closed with an even or odd number of crossings.
The even number of crossings class is equivalent to the non crossing case, while the odd number of crossings class is equivalent to the one crossing case.
As described in \cite{Bonzom:2017:DiagrammaticsColoredSYK} the graphs contributing to $G_{\text{NLO}}$ exist in two types $A$ and $B$, which are themselves separated in two subtypes $\emptyset$ or $\text{not } \emptyset$.
The second subtype always contributes to the first term of the right hand side of equation \eqref{eq:SE-NLO-SYK}, while the type $A,\ \emptyset$ (respectively $B,\ \emptyset$) case accounts for the second and third (resp.\ fourth and fifth) terms of the right hand side of equation \eqref{eq:SE-NLO-SYK}.
These equations can be rewritten using the further defined color space matrix $Q$.
To this aim we first define a matrix in the color space, whose elements $K_{c, c'}(t_1, t_2; t_3, t_4)$ are defined by the equation
\begin{equation}
	K_{c, c'}(t_1, t_2; t_3, t_4) = - g \, (1 - \delta_{c, c'}) \, G_{[0]}(t_1, t_3) G_{[0]}(t_2, t_4) G_{[0]}(t_3, t_4)^{q-2},
\end{equation} for $c, c' \in\{1, \ldots, q\}$.
The analogue of this operator re-appears with slight modifications in the tensor model context as well.
One defines the matrices $Q_0$ and $Q$, whose elements are
\begin{align}
	\label{eq:Q1def-SYK}
	Q_{0, c, c'}(t_1, t_2; t_3, t_4)
		&= \delta_{c, c'} \big( G_{[0]}(t_1, t_3) G_{[0]}(t_2, t_4)- G_{[0]}(t_1, t_4) G_{[0]}(t_2, t_3) \big) \\
	Q_{c, c'}(t_1, t_2; t_3, t_4)
		&= \sum_{n\ge 0} \big[K^n \ast Q_0 \big]_{c, c'}(t_1, t_2; t_3, t_4) \\
		&= \left[(\delta^{\otimes 2} \otimes \mathbbm{1} - K)^{-1} Q_0 \right]_{c, c'}(t_1, t_2; t_3, t_4)
\end{align}
where $\delta^{\otimes 2} = \delta(t_1-t_3) \delta(t_2-t_4)$ and the $\ast$ product here means both matrix and convolution product of the form
\begin{equation}
	\left[K \ast Q_0 \right]_{c, c'}(t_1, t_2; t_3, t_4)
		= \sum_{c''} \int \dd t \dd t' \,
			K_{c, c''}(t_1, t_2; t, t') Q_{0, c'', c'}(t, t'; t_3, t_4),
\end{equation}
and the powers $n$ of $K$ are taken with respect to this product.

Notice here that equation \eqref{eq:Q1def-SYK} is singular if $K$ admits an eigenvector with eigenvalue $1$, which is the case in the large coupling limit.
As explained in the introduction (\cref{sec:introduction}) this signals a spontaneous breaking of the conformal symmetry and for this reason this contribution can be ignored: it is an artifact of the limit which can be handled by including subleading corrections in the coupling constant.
Since the latter break the conformal symmetry any statement about the conformal symmetry assumes that one is considering the large coupling limit with the divergent contribution removed~\cite{Maldacena:2016:CommentsSachdevYeKitaevModel}.

If we consider the 1PI counterpart of $Q$, written as $\Gamma$, we find that it satisfies Schwinger--Dyson-like equations of the form
\begin{equation}
	\label{eq:1PI-4points-SYK}
	\Gamma(t_1, t_2; t_3, t_4)
		= \Gamma_0(t_1, t_2; t_3, t_4) + [\Gamma \ast K](t_1, t_2; t_3, t_4)
\end{equation}
and $\Gamma_0$ writes element-wise
\begin{equation}
	\Gamma_{0, c, c'}(t_1, t_2; t_3, t_4)
		= g \, (1 - \delta_{c, c'}) \delta(t_1-t_3) \delta(t_2-t_4)\, G_{[0]}(t_1, t_2)^{q-2}.
\end{equation}
We can rewrite the equation on $\Sigma_{\text{NLO}}$:
\begin{adjustwidth}{-2cm}{-2cm}
\begin{multline}
	\label{eq:SE-NLO-SYK-reduced}
	\Sigma_{\text{NLO}}(t_1, t_2)
		= \sum_{c \neq c_0}
				\raisebox{-6.0ex}{\includegraphics[scale = 0.8]{images/NLOSESYK1stterm.pdf}}
			+ \sum_{c_1, c_2 \neq c_0}
				\raisebox{-12.0ex}{\includegraphics[scale = 0.55]{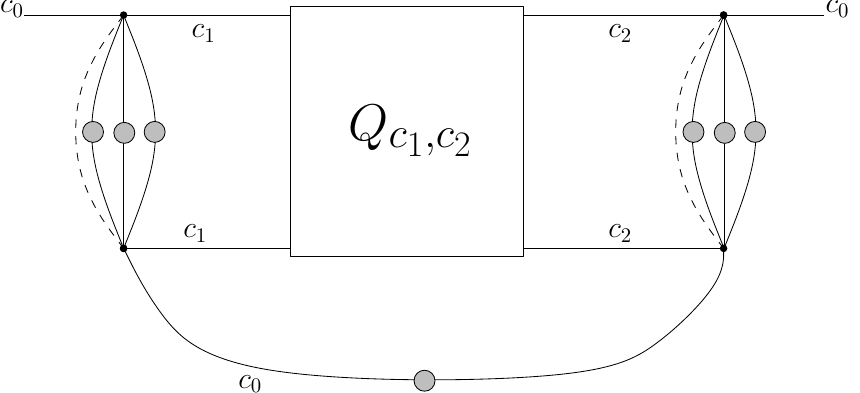}} \\
			+ \sum_{\substack{c_1, c_2 \neq c_0, c_3 \\ c_3 \neq c_0}}
				\raisebox{-12.0ex}{\includegraphics[scale = 0.55]{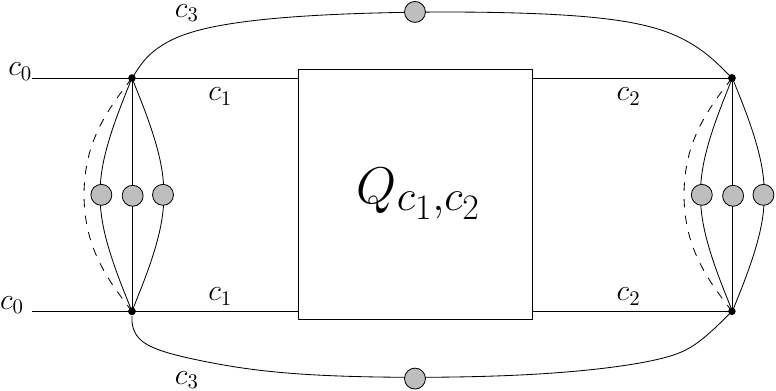}}.
\end{multline}
\end{adjustwidth}
Equation \eqref{eq:SE-NLO-SYK-reduced} rewrites formally as 
\begin{equation}
	\label{eq:SE-NLO-SYK-formal}
	\begin{aligned}
	\Sigma_{\text{NLO}}(t_1, t_2)
		&= (q-1) \, g \; G_{[0]}(t_1, t_2)^{q-2}
				G_{\text{NLO}}(t_1, t_2)
			\\
			& \quad
			+ g^2 \int \dd t \dd t'
				\left[ \sum_{c_1, c_2 \neq c_0} Q_{c_1, c_2}(t_1, t; t_2, t')\right]
				G_{[0]}(t_1, t)^{q-2}
				G_{[0]}(t_2, t')^{q-2}
				G_{[0]}(t, t') 
			\\
			& \quad
			+ g^2 \int \dd t \dd t'
				\left[ \sum_{c_3 \neq c_0} \sum_{c_1, c_2 \neq c_0, f_3} Q_{c_1, c_2}(t_2, t_2; t, t')\right]
				G_{[0]}(t_1, t_2)^{q-3}
				\\
				& \hspace{6cm}
				\times
				G_{[0]}(t_1, t)
				G_{[0]}(t_2, t')
				G_{[0]}(t, t')^{q-2}
		\\
		&:= \Sigma_{\text{NLO}}^{(1)}(t_1, t_2)
			+ \Sigma_{\text{NLO}}^{(2)}(t_1, t_2)
			+ \Sigma_{\text{NLO}}^{(3)}(t_1, t_2).
	\end{aligned}
\end{equation}

We now turn our attention to the NLO $2$-point function.
It can be obtained by rewriting \eqref{eq:SE-to-2pt-exact} in Fourier space:
\begin{equation}
	G_e(\omega)
		= - \frac{1}{i\omega} \left(1 + \frac{\Sigma(\omega)}{i\omega}\right)^{-1}
		= G_f(\omega) \sum_{p\ge 0} \left(- \frac{\Sigma(\omega)}{i\omega} \right)^p.
\end{equation}
Since $G_e(\omega) = \sum_{\ell_m\ge 0} N^{-\ell_m} G_{[\ell_m]}(\omega)$ and $\Sigma(\omega) = \sum_{\ell_m\ge 0}N^{-\ell_m}\Sigma_{[\ell_m]}(\omega)$, we have 
\begin{align}
	G_{\text{NLO}}(\omega)
		&= \left[ G_f(\omega) \sum_{q\ ge 0}
			\left(- \frac{\Sigma_{[0]}(\omega)}{i\omega} \right)^q
			\right]
			\Sigma_{\text{NLO}}(\omega)
			\left[G_f(\omega) \sum_{p \ge 0}
				\left(- \frac{\Sigma_{[0]}(\omega)}{i\omega} \right)^p
				\right]
		\\
		&= \big( G_f(\omega)^{-1} - \Sigma_{[0]}(\omega) \big)^{-1} \Sigma_{\text{NLO}}(\omega)\big( G_f(\omega)^{-1} - \Sigma_{[0]}(\omega) \big)^{-1}
		\\
		&= G_{[0]}(\omega) \Sigma_{\text{NLO}}(\omega) G_{[0]}(\omega).
\end{align}
Transforming back this expression to position space leads to:
\begin{equation}
	\label{eq:convolutionNLO-SYK}
	G_{\text{NLO}}(t_1, t_2)
		= \int \dd t \dd t' \,
			G_{[0]}(t_1, t) \Sigma_{\text{NLO}}(t, t') G_{[0]}(t', t_2).
\end{equation}
Inserting \eqref{eq:SE-NLO-SYK-formal} gives an integral equation on $G_{\text{NLO}}$:
\begin{equation}
	\begin{aligned}
	G_{\text{NLO}}(t_1, t_2)
		= (q - 1) \, g
				&\
				\int \dd t \dd t' \,
				G_{[0]}(t_1, t) G_{[0]}(t, t')^{q-2} G_{\text{NLO}}(t, t') G_{[0]}(t', t_2)
			\\
			+ &\int \dd t \dd t' \,
				G_{[0]}(t_1, t) \big( \Sigma_{\text{NLO}}^{(2)}(t, t') + \Sigma_{\text{NLO}}^{(3)}(t, t') \big) G_{[0]}(t', t_2).
	\end{aligned}
\end{equation} 
This can be simplified further by recognizing the operator $K_{cc}$:
\begin{equation}
	\label{eq:convolutionNLO-inteq}
	\begin{multlined}
	\int \dd t \dd t' \,
			\big[ \delta(t_1 - t) \delta(t_2 - t') - K(t_1, t_2 ; t, t') \big]
			G_{\text{NLO}}(t, t')
		\\
		= \int \dd t \dd t' \,
			G_{[0]}(t_1, t) \big( \Sigma_{\text{NLO}}^{(2)}(t, t') + \Sigma_{\text{NLO}}^{(3)}(t, t') \big) G_{[0]}(t', t_2).
	\end{multlined}
\end{equation} 
Note that $K := \sum_c K_{cc}$ which contains a factor $\sum_c \delta_{cc} = q$.
Defining the inverse of $(1 - K)$ by $L$
\begin{equation}
	\label{eq:inv-L}
	\int \dd t \dd t'
			\big[ \delta(t_1 - t) \delta(t_2 - t') - K(t_1, t_2 ; t, t') \big]
			L(t, t'; t_3, t_4)
		= \delta(t_1 - t_3) \delta(t_2 - t_4),
\end{equation} 
the final expression for $G_{\text{NLO}}$ reads:
\begin{equation}
	\label{eq:GNLO-sol-int}
	G_{\text{NLO}}(t, t')
		= \int \dd t_1 \dd t_2 \dd t_3 \dd t_4 \,
			L(t, t'; t_1, t_2)
			G_{[0]}(t_1, t_3)
			\big( \Sigma_{\text{NLO}}^{(2)}(t_3, t_4) + \Sigma_{\text{NLO}}^{(3)}(t_3, t_4) \big) G_{[0]}(t_4, t_2).
\end{equation} 

We now want to study the scaling dimension of the NLO in the large coupling limit.
The large coupling limit of \eqref{eq:GNLO-sol-int} follows by adding bars on all quantities:\footnotemark{}
\footnotetext{%
	In a previous version, we had assumed incorrectly that the first term in \eqref{eq:SE-NLO-SYK-formal} is subleading with respect to the other terms in this regime.
	Part of the origin of this incorrect statement is a typo in eq.~(51) of~\cite{Kaminski:2014:DoublescalingLimitTensor}: the equation should read $G_{\text{NLO}} \sim g^2 \pd_g G_{\text{LO}}^{q+1}$, whereas the exponent in~\cite{Kaminski:2014:DoublescalingLimitTensor} was $q+2$.
	However, this assumption is in fact not necessary and we can proceed differently.
}%
\begin{equation}
	\label{eq:NLO-SYK-largeg}
	\begin{gathered}
	\bar G_{\text{NLO}}(t, t')
		= \int \dd t_1 \dd t_2 \dd t_3 \dd t_4 \,
			\bar L(t, t'; t_1, t_2)
			\bar G_{[0]}(t_1, t_3)
			\big( \bar \Sigma_{\text{NLO}}^{(2)}(t_3, t_4) + \bar \Sigma_{\text{NLO}}^{(3)}(t_3, t_4) \big) \bar G_{[0]}(t_4, t_2).
	\\
	\int \dd t \dd t'
			\big[ \delta(t_1 - t) \delta(t_2 - t') - \bar K(t_1, t_2 ; t, t') \big]
			\bar L(t, t'; t_3, t_4)
		= \delta(t_1 - t_3) \delta(t_2 - t_4),
	\\
	\bar \Sigma_{\text{NLO}}^{(2)}(t_1, t_2)
		= g^2 \int \dd t \dd t'
			\left[ \sum_{c_1, c_2 \neq c_0} \bar Q_{c_1, c_2}(t_1, t; t_2, t')\right]
			\bar G_{[0]}(t_1, t)^{q-2}
			\bar G_{[0]}(t_2, t')^{q-2}
			\bar G_{[0]}(t, t'),
	\\
	\begin{multlined}
	\bar \Sigma_{\text{NLO}}^{(3)}(t_1, t_2)
		= g^2 \int \dd t \dd t'
				\left[ \sum_{c_3 \neq c_0} \sum_{c_1, c_2 \neq c_0, f_3} \bar Q_{c_1, c_2}(t_2, t_2; t, t')\right]
				\bar G_{[0]}(t_1, t_2)^{q-3}
				\\
				\times
				\bar G_{[0]}(t_1, t)
				\bar G_{[0]}(t_2, t')
				\bar G_{[0]}(t, t')^{q-2}
	\end{multlined}
	\end{gathered}
\end{equation} 
Hence, we need to find the transformation properties of all the objects which appear in these formulas before finding the transformation of $G_{\text{NLO}}$.

To this aim we come back to the equations \eqref{eq:1PI-4points-SYK} and use \eqref{eq:syk-reparam-2pt} repetitively.
These imply that, in the conformal sector, the 1PI counterpart of $Q\rightarrow \bar{Q}$ has scaling dimension $(q-1)/q$.
Indeed, it is easy to check that the terms $\bar{\Gamma}_0$ has $(q-1)/q$ as scaling dimensions:
\begin{equation}
	\label{eq:Gamma0scaling}
	\bar{\Gamma}_{0, c, c'}(\sigma_1, \sigma_2; \sigma_3, \sigma_4)
		= \frac{\bar{\Gamma}_{0, c, c'}(t_1, t_2; t_3, t_4)}{|f'(t_1) f'(t_2) f'(t_3) f'(t_4)|^{(q-1)/q}},
\end{equation}
for $\sigma_i = f(t_i)$, $i = 1, \ldots, 4$.
This follows from:
\begin{align*}
	\bar{\Gamma}_{0, c, c'}(\sigma_1, \sigma_2; \sigma_3, \sigma_4)
		&= (q-2) g\, (1 - \delta_{c, c'}) \,
			\frac{\delta(t_1 - t_3) \delta(t_2 - t_4)}{|f'(t_3) f'(t_4)|}
			\frac{\bar{G}_{[0]}(t_1, t_2)^{q-2}}{|f'(t_1) f'(t_2)|^{(q-2)/q}}
		\\
		&= \frac{|f'(t_3) f'(t_4)|^{1/q}}{|f'(t_1) f'(t_2)|^{1/q}} \,
			\frac{\bar{\Gamma}_{0, c, c'}(t_1, t_2; t_3, t_4)}{|f'(t_3) f'(t_4)| \, |f'(t_1) f'(t_2)|^{(q-2)/q}},
\end{align*}
using \eqref{eq:syk-reparam-2pt}.

Let us consider the terms of the form $\left[ \bar{\Gamma} \ast K\right](t_1, t_2; t_3, t_4)$.
From the definition of $K$ and \eqref{eq:syk-reparam-2pt} we find:
\begin{equation}
	\label{eq:Kscaling}
	\bar{K}_{c, c'}(\sigma_1, \sigma_2; \sigma_3, \sigma_4)
		= \frac{ |f'(t_3) f'(t_4)|^{1/q-1}}{|f'(t_1) f'(t_2)|^{1/q}} \,
			\bar{K}_{c, c'}(t_1, t_2; t_3, t_4).
\end{equation}
Therefore, using equations \eqref{eq:Gamma0scaling} and \eqref{eq:Kscaling} it is simple to check that if $\bar{\Gamma}(t_1, t_2; t_3, t_4)$ is a solution of \eqref{eq:1PI-4points-SYK} in the conformal sector, then the equation satisfied by $\bar{\Gamma}(\sigma_1, \sigma_2; \sigma_3, \sigma_4)$ transforms into \eqref{eq:1PI-4points-SYK} for $\sigma_i = f(t_i)$ provided that
\begin{equation}
	\bar{\Gamma}(t_1, t_2; t_3, t_4)
		= |f'(t_1) f'(t_2) f'(t_3) f'(t_4)|^{1-1/q} \,
			\bar{\Gamma}(\sigma_1, \sigma_2; \sigma_3, \sigma_4).
\end{equation}
One is then interested in the scaling dimension of $\bar Q$.
We have
\begin{equation}
	\begin{aligned}
		\bar{Q}(t_1, t_2; t_3, t_4)
			= \bar{Q}_0(t_1, t_2; t_3, t_4)
			+ \int \dd t \dd t' \dd \tau \dd \tau'\, \Big(&
				\big[\bar{G}_{[0]}(\tau, t_3) \bar{G}_{[0]}(\tau', t_4) 
					- \bar{G}_{[0]}(\tau, t_4) \bar{G}_{[0]}(\tau', t_3) \big]
					\\
				&\times \bar{G}_{[0]}(t_1, t) \bar{G}_{[0]}(t_2, t') \bar{\Gamma}(t, t';\tau, \tau')
				\Big),
	\end{aligned}
\end{equation}
where the integration is done element-wise.
From this last equality, one shows that the scaling dimension of $\bar{Q}(t_1, t_2; t_3, t_4)$ is $1/q$ by using the scaling properties of $\bar{G}_{[0]}$ as well as the ones of $\bar{\Gamma}$.
Then, as we know that $\bar{Q}$ has scaling dimension $1/q$, a simple computation shows that $\bar{\Sigma}_{\text{NLO}}^{(2)}$ and $\bar{\Sigma}_{\text{NLO}}^{(3)}$ have scaling dimension $\frac{q-1}{q}$.

Knowing the transformation of $\bar K$, we can study the one of $\bar L = (1 - \bar K)^{-1}$.
We start from the definition \eqref{eq:inv-L} and perform a reparametrization of both sides:
\begin{align*}
	\delta(\sigma_1 - \sigma_3) \delta(\sigma_2 - \sigma_4)
		&= \frac{\delta(t_1 - t_3) \delta(t_2 - t_4)}{|f'(t_3) f'(t_4)|}
	\\
	&= \int \dd \sigma \dd \sigma'
		\big[ \delta(\sigma_1 - \sigma) \delta(\sigma_2 - \sigma') - \bar K(\sigma_1, \sigma_2 ; \sigma, \sigma') \big]
			\bar L(\sigma, \sigma'; \sigma_3, \sigma_4)
	\\
	&
	\begin{multlined}
	= \int \dd t \dd t' \, |f'(t) f'(t')|
		\left[
			\frac{\delta(t_1 - t) \delta(t_2 - t')}{|f'(t) f'(t')|}
			- \frac{ |f'(t) f'(t')|^{1/q-1}}{|f'(t_1) f'(t_2)|^{1/q}} \,
			\bar{K}(t_1, t_2; t, t')
			\right]
			\\
			\times \bar L(\sigma, \sigma'; \sigma_3, \sigma_4)
	\end{multlined}
	\\
	&= \int \dd t \dd t' \,
		\frac{ |f'(t) f'(t')|^{1/q}}{|f'(t_1) f'(t_2)|^{1/q}}
		\big[ \delta(t_1 - t) \delta(t_2 - t') - \bar{K}(t_1, t_2; t, t') \big]
			\bar L(\sigma, \sigma'; \sigma_3, \sigma_4)
\end{align*}
for $\sigma, \sigma' = f(t), f(t')$.
Hence, consistency  between both sides leads to the transformation
\begin{equation}
	\bar L(\sigma_1, \sigma_2; \sigma_3, \sigma_4)
		= \frac{ |f'(t_3) f'(t_4)|^{1/q-1}}{|f'(t_1) f'(t_2)|^{1/q}} \,
			\bar{L}(t_1, t_2; t_3, t_4).
\end{equation} 

Thanks to these different relations, we can deduce the scaling dimension of $\bar{G}_{\text{NLO}}$:
\begin{equation}
	\bar{G}_{\text{NLO}}(\sigma_1, \sigma_2)
		= \frac{\bar{G}_{\text{NLO}}(t_1, t_2)}{|f'(t_1) f'(t_2)|^{1/q}}.
\end{equation} 
This can in turn be used to determine the scaling of the self-energy:
\begin{equation}
	\bar{\Sigma}_{\text{NLO}}(\sigma, \sigma')
		= |f'(t) f'(t')|^{1/q-1} \,
			\bar{\Sigma}_{\text{NLO}}(t, t').
\end{equation} 
As a consistency check, every term of strong coupling limit of \eqref{eq:SE-NLO-SYK-formal} transforms in the same way.\footnotemark{}
\footnotetext{%
	Another derivation would have been to assume that each term must transform and to make the ansatz that $G_{\text{NLO}}$ scales with a power-law.
	This would determine the power of the transformation, and self-consistency of the ansatz can be checked by plugging the result in \eqref{eq:convolutionNLO-SYK}.
}%
As a consequence the scaling dimension of $\bar{G}_{\text{NLO}}$ is $1/q$ in the conformal sector.
This is the same scaling dimension than $\bar{G}_{[0]}$, thus the conformal symmetry is not altered at NLO in $N$ in the large coupling limit.

\section{Real and complex colored tensor SYK models}
\label{sec:syk-colored-tensor}

\subsection{The models}

In this part, we consider one dimensional fermionic quantum field tensor models.
The first one is built out of real fermionic fields, while the second one is built from complex fermionic fields.
Each field carries a color index $c$ plus $D$ additional indices denoting the component of the tensor.\footnotemark{}
\footnotetext{%
	In this section, $D$ plays the same role as $q$.
	We keep the notations different to emphasize which model is under study.
}%

\medskip

The real model is the Gurau--Witten model introduced in \cite{Witten:2016:SYKLikeModelDisorder}.
Its partition function writes,
\begin{equation}
	Z^{\mathbb{R}}_{N, \lambda}
		= \int \prod_{c = 0}^D \mathcal{D}\psi^c\, \e^{- \int \dd t\, L[\psi]}
\end{equation} 
where
\begin{equation}
	L[\psi]
		= \frac{1}{2} \sum_{c = 0}^D \sum_{n_c} \psi^c_{n_c} \partial_t \psi^c_{n_c}
		+ \I^{(D + 1)/2} \frac{\lambda}{N^{D(D-1)/4}} \sum_{n} \prod_{c = 0}^D \psi^c_{n_c}.
\end{equation} 
It is also convenient to define
\begin{equation}
	g = \lambda^2.
\end{equation} 
We now need to explain several points.
Let us first start with the notations.
As explained above the fermionic fields are tensors.
As such they are $D$-fundamentals of $\group{O}(N)$.
The tensors carry a color index $c$ which runs from $0$ to $D$.
This means we have a family of $D + 1$ fermionic tensor fields $\{\psi^c\}_{c = 0}^D$.
Since each $\psi^c$ is a tensor, its components write $\psi^c_{n^{c0} \cdots n^{cD}}$ for $n^{cj}$ ranging from $1$ to $N$.
We call $N$ the size of the tensor, each field $\psi^c$ has $N^D$ components.
Then $\sum_{n_c}\psi^c_{n_c}\partial_t\psi^c_{n_c}$ means
\begin{equation}
	\sum_{n_c} \psi^c_{n_c} \partial_t \psi^c_{n_c} := \sum_{n^{c0} \cdots n^{cD} \ge 1} \psi^c_{n^{c0} \cdots n^{cD}} \partial_t \psi^c_{n^{c0} \cdots n^{cD}}.
\end{equation}
The interaction term notation $\sum_{n} \prod_{c = 0}^d\psi^c_{n_c}$ contains $\sum_n$ which is a shorthand for the constraint that $n_c = (n^{c (c-1)} \cdots n^{c0}n^{cD} \cdots n^{c (c + 1)})$ and that the indices are constrained to $n^{kl} = n^{lk}$.

\medskip

In this model the free scalar two-point function $G_f$ is 
\begin{equation}
	G_f(t_1, t_2) = \frac{1}{N^D} \, \Big\langle \sum_{n_i} T \psi^c_{n_i}(t_1) \psi^c_{n_i}(t_2) \Big\rangle_0 = \frac{1}{2}\, \sign(t_1-t_2), 
\end{equation}
its Fourier transform writes,
\begin{equation}
	G_f(\omega) = - \frac{1}{i\omega}.
\end{equation} 
Meanwhile we have 
\begin{equation}
	\Big\langle T \psi^{c'}_{n_{c'}}(t_1) \psi^c_{n_c}(t_2) \Big\rangle_0 = \left(\delta_{cc'} \prod_{c_1\neq c} \delta_{n^{c'c_1} n^{cc_1}}\right) G_f(t_1, t_2).
\end{equation}
The exact two-point function $G_e$ on the other hand satisfies the same type of relation 
\begin{align}
	G_e(t_1, t_2) &= \frac{1}{N^D} \, \Big\langle \sum_{n_i} T \psi^c_{n_i}(t_1)\psi^c_{n_i}(t_2) \Big\rangle \\
	\Big\langle T \psi^{c'}_{n_{c'}}(t_1) \psi^c_{n_c}(t_2) \Big\rangle &= \left(\delta_{cc'} \prod_{c_1\neq c}\delta_{n^{c'c_1}n^{cc_1}}\right) G_e(t_1, t_2).
\end{align}

\medskip

The complex model is very similar to the real one.
It is constructed out of $2(D + 1)$ complex fermionic tensor fields $\psi^{c}_{n_c}(t), \bar{\psi}^{c}_{n_c}(t)$.
$c \in [0..D]$ is the color of the tensor, and each subscript $n_c$ is an abbreviation of the form $n_i = \{ n^{cc-1}, \ldots, n^{c0}, n^{cD}, \ldots, n^{cc + 1}\}$, where each $n^{ij}\in [1..N]$ for some $N$, again the size of the tensors.
The corresponding partition function is 
\begin{equation}
	Z_{N, \lambda, \bar{\lambda}}^{\mathbb{C}} = \int \prod_{i = 0}^D \mathcal{D} \psi^i \mathcal{D} \bar{\psi^i}\, \e^{ \int \dd t\, L[\psi]}.
\end{equation}
where
\begin{equation}
	\begin{aligned}
		L[\psi] = \sum_{c = 0}^D \sum_{n_c} \bar{\psi}^c_{n_c} \partial_t \psi^c_{n_c}
			&+ \I^{(D + 1)/2} \frac{\lambda}{N^{D(D-1)/4}} \sum_{n} \prod_{c = 0}^D \psi^c_{n_c} \\
			&+ \I^{(D + 1)/2} \frac{\bar{\lambda}}{N^{D(D-1)/4}} \sum_{n} \prod_{c = 0}^D \bar{\psi}^c_{n_c}.
	\end{aligned}
\end{equation} 
The definition of the sum in the interaction term is the same than in the real case.
Each fermion field is a $d$-fundamental of $\group{U}(N)$ and we will make use of the notation
\begin{equation}
	g = \lambda \bar\lambda.
\end{equation} 

The two-point functions are defined in similar ways.
The free two-point function satisfies
\begin{align}
	G_f(t_1, t_2) &= \frac{1}{N^D} \, \Big\langle \sum_{n_i} T \bar{\psi}^c_{n_i}(t_1) \psi^c_{n_i}(t_2) \Big\rangle_0
		= \operatorname{\sign}(t_2 - t_1), \\
	\Big\langle T \bar{\psi}^{c'}_{n_{c'}}(t_1) \psi^c_{n_c}(t_2) \Big\rangle_0 &= \left(\delta_{cc'} \prod_{c_1 \neq c}\delta_{n^{c'c_1} n^{cc_1}} \right) G_f(t_1, t_2),
\end{align}
while the exact two-point function satisfies
\begin{align}
	G_e(t_1, t_2) &= \frac{1}{N^D} \, \Big\langle \sum_{n_i} T \bar{\psi}^c_{n_i}(t_1) \psi^c_{n_i}(t_2) \Big\rangle \\
	\Big\langle T \bar{\psi}^{c'}_{n_{c'}}(t_1) \psi^c_{n_c}(t_2) \Big\rangle &= \left(\delta_{cc'} \prod_{c_1 \neq c} \delta_{n^{c'c_1}n^{cc_1}} \right) G_e(t_1, t_2).
\end{align}
We make a slight abuse of notations here as we use the same notations for both the complex and real case.
In fact this is to avoid introducing too many notations.

\medskip

We now describe the Feynman graphs of these models.
The Feynman graphs have the following properties:
\begin{itemize}
	\item The vertices are $(D + 1)$-valent.

	\item Edges carry a color index $c$ ranging from $0$ to $D$ in such a way that no two adjacent edges have the same color index.
	
	\item the faces of the graphs are the bicolored edge cycles.
	
	\item In the complex case, the graphs are bipartite.
\end{itemize}
The free energy of these models has a $1/N$ expansion driven by the degree $\varpi$,
\begin{equation}
	F_{N, \lambda, \bar{\lambda}} = \log Z_{N, \lambda, \bar{\lambda}}
		= \sum_{\varpi\ge 0}N^{D - \frac{2}{(D-1)!} \varpi} F_{[\varpi]}(\lambda, \bar{\lambda}),
\end{equation}
where the degree $\varpi$ of a graph $\mathcal{G}$ is computed of the genera of its jackets, see \cite{Gurau:2011:1NExpansionColored}, its amplitude is then $A(\mathcal{G}) = N^{D - \frac{2}{(D-1)!} \varpi(\mathcal{G})} a(\mathcal{G})$ where $a(\mathcal{G})$ is a reduced amplitude that depends on integral over positions and the coupling constants but not on $N$.
The main difference between the complex and real case is that, \emph{a priori}, the degree in the complex case is an integer because all jackets are ribbon graphs representing surfaces, while in the real case, non-orientable two manifolds can appear among the jackets and thus turn the degree into an half-integer.
However, it is easy to show that the degree is an integer in both cases.

The fixed degree free energies $F^{[\varpi]}_{\lambda, \bar{\lambda}}$ can be computed by summing\footnotemark{} the amplitudes of all vacuum connected Feynman graphs of degree $\varpi$.
\footnotetext{%
	Actually one should be more precise here.
	By summing all the amplitudes one gets the perturbative free energies.
	However these free energies are likely to have a finite radius of convergence in the coupling constant, and thus be defined only in a disc type domain around $\lambda^2 = 0$.
	As a consequence, if one is interested in large coupling physics one should find the (possibly many) analytic continuations of these perturbative free energies.
	Another way to consider the large coupling case is to find functional equations for the free energies and solve them in the large coupling regime.
	These functional equations can sometimes be found using only perturbative arguments, this is exactly what is done for the leading order two-point function.
}%

These considerations imply that the two-point function also has a $1/N$ expansion.
This expansion writes in both the real and complex cases
\begin{equation}
	\label{eq:Ge-expansion}
	G_e(t_1, t_2) = \sum_{\varpi\ge 0} N^{- \frac{2}{(D-1)!} \varpi} G_{[\varpi]}(t_1, t_2).
\end{equation}

\subsection{The Leading Order}

The leading order of the $1/N$ expansion, $\varpi = 0$, is described by melon diagrams.
They are graphs of degree $0$, meaning that all jackets are planar.
Thanks to the structural properties of the melonic graphs, it is easy to infer the equation satisfied by the LO $2$-point function.
Indeed, one has the usual relation between the self-energy $\Sigma$ and the exact two-point function:
\begin{equation}\label{eq:2pointtoself}
	G_e(t_1, t_2) = \big( G_f(t_1, t_2)^{-1} - \Sigma(t_1, t_2) \big)^{-1},
\end{equation}
where the inverse is taken with respect to the matrix-like/convolution product.
Recalling that one writes $g = \lambda^2$ in the real case or $g = \lambda \bar \lambda$ in the complex case, one deduces that at leading order, 
\begin{equation}
	G_{[0]}(t_1, t_2) = G_f(t_1, t_2) + g \int \dd t \dd t' \, G_f(t_1, t) \Sigma_{[0]}(t, t') G_{[0]}(t', t_2),
\end{equation}
where $G_f$ is the free field two-point function.
Then the structural properties of melonic graphs implies that 
\begin{equation}
	\Sigma_{[0]}(t, t') = G_{[0]}(t, t')^D.
\end{equation} 
This equation can be reduced in the infrared/large coupling limit.
Indeed, in the Fourier space this equation rewrites
\begin{equation}
	- \I \omega \, G_{[0]}(\omega) = 1 + g \, \Sigma_{[0]}(\omega) G_{[0]}(\omega),
\end{equation}
where we introduced the notation $\Sigma_{[0]}(\omega)$ for the Fourier transform of the self-energy at leading order in $N$.
In the infrared limit, the left hand side drops.
If we introduce $\bar{G}_{[0]}$ the infrared/large coupling limit of $G_{[0]}$ then $\bar{G}_{[0]}$ satisfies the equation, 
\begin{equation}
	0 = 1 + g \, \bar{\Sigma}_{[0]}(\omega) \bar{G}_{[0]}(\omega),
\end{equation}
which rewrites in position space as,
\begin{equation}
	\label{eq:largegLOequation}
	g \int \dd t' \, \bar{G}^D_{[0]}(t_1, t') \bar{G}_{[0]}(t', t_2) = - \delta(t_1-t_2).
\end{equation}
The explicit solution in this limit is given by~\cite{Maldacena:2016:CommentsSachdevYeKitaevModel}
\begin{equation}
	\label{eq:largegLOsolution}
	\bar{G}_{[0]}(t_1, t_2) = \left(\frac{(D-1) \tan(\pi/(D + 1))}{2\pi(D + 1)g} \right)^{1/(D + 1)} \frac{\sign(t_1-t_2)}{|t_1 - t_2|^{2/(D + 1)}}.
\end{equation}

\medskip

Coming back to the equation \eqref{eq:largegLOequation} satisfied by $\bar{G}_{[0]}$, one can show that if $\bar{G}_{[0]}(t_1, t_2)$ is a solution, then, $\bar{G}_{[0]}(\sigma_1, \sigma_2)$, where $\sigma_{1, 2} = f(t_{1, 2})$, is a solution as well, provided that $\bar{G}_{[0]}(t_1, t_2) = |\partial_{t_1} f(t_1) \partial_{t_2} f(t_2)|^{\frac{1}{D + 1}} \bar{G}_{[0]}(\sigma_1, \sigma_2)$.
$\frac{1}{D + 1}$ is the scaling dimension of $\bar{G}_{[0]}$.

\subsection{Next-to-Leading Order two-point function}

We want to study the Next-to-Leading Order of the real and complex colored tensor models.
The goal is to check whether or not these models display the conformal symmetry property at large coupling.
In particular to check if it is true or not, we need to compute the scaling dimension of the two-point function at NLO.
We then study the two-point function at NLO.

\medskip

As is seen in \cite{Bonzom:2017:DiagrammaticsColoredSYK}, the NLO of the real and complex model are described by the same family of Feynman graphs.
This means that non bipartite graphs do not appear at NLO.
This is a specificity of the NLO that is not recovered at all orders.
The complex cases have been investigated in the zero dimensional bosonic tensor model case in \cite{Kaminski:2014:DoublescalingLimitTensor}.
Following \cite{Kaminski:2014:DoublescalingLimitTensor}, it is possible to show that the value of the degree at NLO is 
\begin{equation}
	\varpi_{\text{NLO}} = \frac{(D-1)!}{2}\, (D-2).
\end{equation}
The NLO 1PI self-energy and two-point functions are defined by
\begin{equation}
	G_{\text{NLO}}(t_1, t_2) := G_{\left[ \frac{(D-1)!}{2}(D-2)\right]}(t_1, t_2), \qquad
	\Sigma_{\text{NLO}}(t_1, t_2) := \Sigma_{\left[ \frac{(D-1)!}{2}(D-2)\right]}(t_1, t_2).
\end{equation}
The functional equation for the 1PI self-energy writes graphically
\begin{adjustwidth}{-0.5cm}{-0.5cm}
\begin{multline}
	\label{eq:1PINLOgraphic}
	\Sigma_{\text{NLO}}(t_1, t_2) = \sum_{c\neq c_0}
			\raisebox{-5.0ex}{\includegraphics[scale = 0.8]{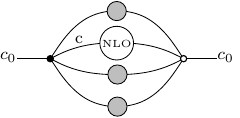}}
		+ \sum_{c_1\neq c_0}
			\raisebox{-12.0ex}{\includegraphics[scale = 0.6]{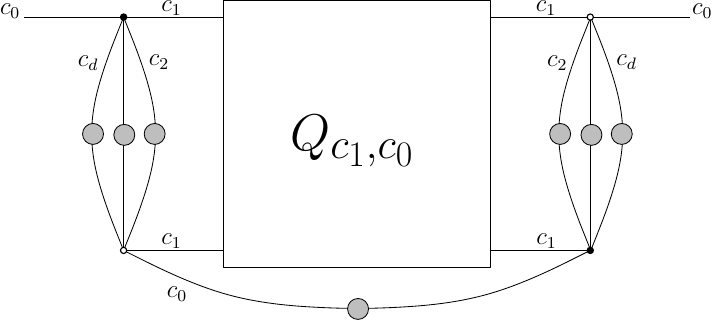}} \\
		+ \sum_{\substack{\textrm{pairs }\{c_2, c_1\}\\ c_2, c_1\neq c_0\\ c_1\neq c_2}}
			\raisebox{-12.0ex}{\includegraphics[scale = 0.6]{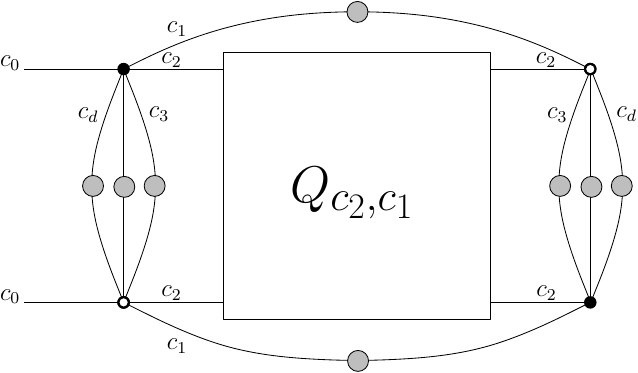}}.
\end{multline}
\end{adjustwidth}
The edges with grey disk insertion represent leading order two-point functions.
The box represents one of the $Q_{c_i, c_j},\ c_i \neq c_j, $ which are the sum of ladder graphs of even length with ingoing/outgoing color $c_i$ and transmitted colors both $c_i$ and $c_j$ (unbroken chains in the language of \cite{Gurau:2017:Complete1NExpansion}), so to say we have 
\begin{adjustwidth}{-2cm}{-2cm}
\begin{multline}
	Q_{c_i, c_j}(t_1, t_2;\tau_1, \tau_2) = \raisebox{-8.5ex}{\includegraphics[scale = 0.65]{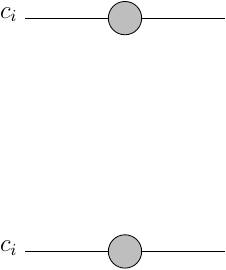}}
		+ \raisebox{-8.5ex}{\includegraphics[scale = 0.65]{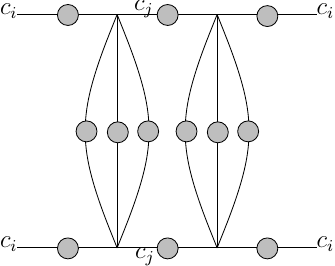}} 
		+ \raisebox{-8.5ex}{\includegraphics[scale = 0.65]{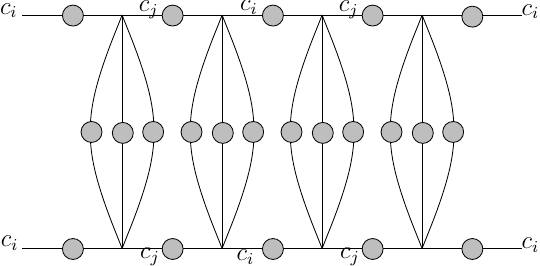}} 
		+ \cdots \\
\end{multline}
\end{adjustwidth}
We have the helpful property that $Q_{c_i, c_j}(t_1, t_2;\tau_1, \tau_2) = Q_{c_n, c_m}(t_1, t_2;\tau_1, \tau_2)$ for any choice of $c_i, c_j$ and $c_n, c_m$.
Then we call $Q_{c_i, c_j}(t_1, t_2; \tau_1, \tau_2) = Q(t_1, t_2; \tau_1, \tau_2)$.
Equation \eqref{eq:1PINLOgraphic} rewrites formally
\begin{equation}
	\label{eq:1PINLO}
	\begin{aligned}
		\Sigma_{\text{NLO}}(t_1, t_2) =\ & D g \, G_{[0]}(t_1, t_2)^{D-1} G_{\text{NLO}}(t_1, t_2) \\
			&+ D g^2 \int \dd t \dd t' \, G_{[0]}(t_1, t)^{D-1} G_{[0]}(t', t_2)^{D-1} G_{[0]}(t, t') Q(t_1, t;t_2, t') \\
			&
			\begin{aligned}
				+ \ \frac{D(D-1)}{2} \, g^2 \int \dd t \dd t' \, \Big(&
					G_{[0]}(t_1, t_2)^{D-2} G_{[0]}(t_1, t) G_{[0]}(t_2, t') \\ &\times G_{[0]}(t, t')^{D-1} Q(t_1, t_2; t, t') \Big).
			\end{aligned}
	\end{aligned}
\end{equation}
We also have
\begin{equation}\label{eq:convolutionNLO}
	G_{\text{NLO}}(t_1, t_2) = \int \dd t \dd t' G_{[0]}(t_1, t) \Sigma_{\text{NLO}}(t, t')G_{[0]}(t', t_2).
\end{equation}
Indeed, from the relation \eqref{eq:2pointtoself}, we have in Fourier space,
\begin{align}
	G_e(\omega) &= - \frac{1}{\I \omega} \left(1 + \frac{\Sigma(\omega)}{\I \omega}\right)^{-1}\\
		&= G_f(\omega) \sum_{p\ge 0}\left(- \frac{\Sigma(\omega)}{\I \omega} \right)^p.
\end{align}
Therefore, using the expansion \eqref{eq:Ge-expansion} for $G_e$ and the fact that the self-energy can similarly be expanded
\begin{equation}
	\Sigma(\omega) = \sum_{\varpi\ge 0} N^{- \frac{2}{(D-1)!} \varpi} \Sigma_{[\varpi]}(\omega)
\end{equation} 
we have
\begin{align}
	G_{\text{NLO}}(\omega) &= \left[G_f(\omega) \sum_{q \ge 0} \left(- \frac{\Sigma_{[0]}(\omega)}{\I \omega} \right)^q \right] \Sigma_{\text{NLO}}(\omega) \left[G_f(\omega) \sum_{p \ge 0} \left(- \frac{\Sigma_{[0]}(\omega)}{\I \omega} \right)^p \right] \\
	&= \big( G_f(\omega)^{-1} - \Sigma_{[0]}(\omega) \big)^{-1} \Sigma_{\text{NLO}}(\omega) \big( G_f(\omega)^{-1} - \Sigma_{[0]}(\omega) \big)^{-1} \\
	&= G_{[0]}(\omega) \Sigma_{\text{NLO}}(\omega)G_{[0]}(\omega),
\end{align} which when written in position space leads to equation \eqref{eq:convolutionNLO}.

\medskip

We now take care of $Q(t_1, t_2; t_3, t_4)$ which appears in equation \eqref{eq:1PINLO}.
$Q(t_1, t_2; t_3, t_4)$ can be constructed from $Q_0(t_1, t_2; t_3, t_4)$ using the operator $K$ graphically defined below,
\begin{equation}
	K(t_1, t_2; t_3, t_4) = \raisebox{-8.5ex}{\includegraphics[scale = 0.65]{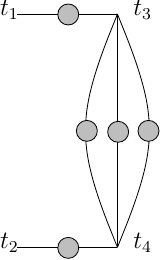}}.
\end{equation}
$Q_0(t_1, t_2; t_3, t_4)$ writes
\begin{equation}
	Q_0(t_1, t_2; t_3, t_4) = G_{[0]}(t_1, t_3) G_{[0]}(t_2, t_4).
\end{equation}
The operator $K$ formally writes as
\begin{equation}
	\label{eq:operator-K}
	K(t_1, t_2, t_3, t_4) = - g \, G_{[0]}(t_1, t_3) G_{[0]}(t_2, t_4) G_{[0]}(t_3, t_4)^{D-1}.
\end{equation} 
We have 
\begin{equation}
	Q(t_1, t_2; t_3, t_4) = \sum_{n\ge 0} K^{2n}(t_1, t_2;t, t') \ast Q_0(t, t'; t_3, t_4)
		= \big( \delta^{\otimes 2} - K \ast K \big)^{-1} \ast Q_0
\end{equation}
where the (even) powers of $K$ are taken with respect to the convolution product.
Again $Q$ is not defined if $K$ possesses eigenvalues $\pm 1$: as explained in \cref{sec:introduction,sec:syk-disorder} we restrict our discussion to the non-divergent part of $Q$.

\medskip

The same argument than in the preceding SYK case applies here and at large $g$: since all terms in the RHS of equation \eqref{eq:1PINLO} scale in the same way and because we only want to find the scaling, we can focus on the simplest two terms:\footnotemark{}
\footnotetext{%
	Keeping several terms help to check the consistency of our computations.
	But, this should not be seen as an approximation because, according to the discussion in \Cref{sec:syk-disorder}, this truncation is not consistent (except to find the scaling).
}%
\begin{align}
	\bar{G}_{\text{NLO}}(t_1, t_2)
		&= \int \dd\tau \dd\tau' \, \bar{G}_{[0]}(t_1, \tau) \bar{\Sigma}_{\text{NLO}}(\tau, \tau') \bar{G}_{[0]}(\tau', t_2) \\
		&
		\begin{aligned}
			\sim \ & D g^2 \int \dd\tau \dd\tau' \dd t \dd t' \, \bar{G}_{[0]}(t_1, \tau)\bar{G}^{D-1}_{[0]}(\tau, t) \bar{G}^{D-1}_{[0]}(t', \tau') \bar{G}_{[0]}(t, t') \bar Q(\tau, t;\tau', t') \bar{G}_{[0]}(\tau', t_2)\\
			&
			\begin{aligned}
				+ \ \frac{D(D-1)}{2} \, g^2 \int \dd\tau \dd\tau' \dd t \dd t' \, \Big(&
					\bar{G}_{[0]}(t_1, \tau) \bar{G}^{D-2}_{[0]}(\tau, \tau') \bar{G}_{[0]}(\tau, t) \bar{G}_{[0]}(\tau', t') \\
					&\times \bar{G}_{[0]}(t, t')^{D-1} \bar{Q}(\tau, \tau';t, t') \bar{G}_{[0]}(\tau', t_2)
					\Big).
			\end{aligned}
		\end{aligned}
\end{align}
In order to get the conformal scaling of $\bar{G}_{\text{NLO}}$ we need to understand how the conformal limit of $\bar{Q}(t_1, t_2; t_3, t_4)$ behaves.
To do so we reduce $\bar Q$ to its 1PI connected counterpart $\bar \Gamma(t_1, t_2; t_3, t_4)$.
We have that $\bar{Q} = \bar{G}^{\otimes 2} + \bar{G}^{\otimes 2} \ast \bar \Gamma \ast \bar{G}^{\otimes 2}$.
More precisely,
\begin{equation}
	\label{eq:definitionQ}
	\begin{aligned}
		\bar{Q}(t_1, t_2; t_3, t_4) = \bar{Q}_0(t_1, t_2; t_3, t_4)
			+ \int \dd t \dd t' \dd\tau \dd\tau' \, \Big(&
				\bar{G}_{[0]}(t_1, t) \bar{G}_{[0]}(t_2, t') \bar \Gamma(t, t';\tau, \tau') \\
				&\times \bar{G}_{[0]}(\tau, t_3) \bar{G}_{[0]}(\tau', t_4) \Big).
	\end{aligned}
\end{equation}
The scaling dimension of $\bar{Q}_0$ is $\frac{1}{D + 1}$ as $\bar{Q}_0$ writes solely in terms of $\bar{G}_{[0]}$.

$\bar \Gamma$ satisfies the following Schwinger--Dyson equation 
\begin{equation}\label{eq:SD4pt1pi}
	\bar \Gamma(t, t'; \tau, \tau') = \Gamma_0(t, t'; \tau, \tau') + \int \dd\eta \dd\eta' \dd\omega \dd\omega' \, \bar \Gamma(t, t';\eta, \eta')\bar{K}(\eta, \eta';\omega, \omega')\bar{K}(\omega, \omega';\tau, \tau'),
\end{equation}
where
\begin{equation}
	\bar \Gamma_0(t, t';\tau, \tau')
		= \raisebox{-8.5ex}{\includegraphics[scale = 0.6]{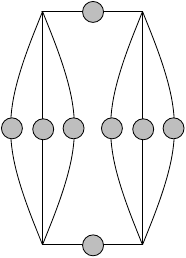}}
		= g^2 \, \bar{G}_{[0]}(t, t')^{D-1} \bar{G}_{[0]}(t, \tau) \bar{G}_{[0]}(t', \tau') \bar{G}_{[0]}(\tau, \tau')^{D-1}
\end{equation}
and we do not display the colors of the edges as the dependence in the times is not sensible to it.
From equation \eqref{eq:SD4pt1pi} we can deduce the scaling of $\bar \Gamma$.
First notice that 
\begin{equation}
	\bar \Gamma_0(\sigma, \sigma'; \zeta, \zeta') = \frac{\bar \Gamma_0(t, t';\tau, \tau')}{|f'(t) f'(t') f'(\tau) f'(\tau')|^{\frac{D}{D + 1}}}
\end{equation} for $\sigma, \sigma', \zeta, \zeta' = f(t), f(t'), f(\tau), f(\tau')$.
This is obtained from the scaling of $\bar{G}_{[0]}$.
Using the explicit expression for $\bar{K}$ we also deduce that 
\begin{multline}
	\int \dd\beta \dd\beta' \, \bar{K}(\sigma, \sigma'; \beta, \beta') \bar{K}(\beta, \beta'; \zeta, \zeta') \\
	\begin{aligned}
		&= \int |f'(\omega) f'(\omega')| \dd\omega \dd\omega' \frac{\bar{K}(\eta, \eta'; \omega, \omega') \bar{K}(\omega, \omega'; \tau, \tau')}{|f'(\eta) f'(\eta')|^{\frac{1}{D + 1}} |f'(\omega) f'(\omega')| |f'(\tau) f'(\tau')|^{\frac{D}{D + 1}}} \\ 
		&= \int \dd\omega \dd\omega' \frac{\bar{K}(\eta, \eta';\omega, \omega') \bar{K}(\omega, \omega'; \tau, \tau')}{|f'(\eta) f'(\eta')|^{\frac{1}{D + 1}} |f'(\tau) f'(\tau')|^{\frac{D}{D + 1}}} 
	\end{aligned}
\end{multline}
where we have set $\sigma, \sigma', \zeta, \zeta'$ as before and $\beta, \beta' = f(\omega), f(\omega')$.
Consequently the scaling dimension of $\bar \Gamma$ is $\frac{D}{D + 1}$.
Indeed if $\bar \Gamma(\sigma, \sigma', \zeta, \zeta')$ is a solution of \eqref{eq:SD4pt1pi}, then the function $\bar \Gamma'(t, t'; \tau, \tau') = |f'(t) f'(t') f'(\tau) f'(\tau')|^{\frac{D}{D + 1}} \bar \Gamma(\sigma, \sigma'; \tau, \tau')$ with $\sigma, \sigma', \zeta, \zeta' = f(t), f(t'), f(\tau), f(\tau')$ is also a solution.

\medskip

We now turn to the scaling dimension of $\bar{Q}$ in the conformal sector.
We recall its expression in terms of $\bar \Gamma$
\begin{equation}
	\label{eq:Qexpression}
	\begin{aligned}
		\bar{Q}(\sigma_1, \sigma_2;\sigma_3, \sigma_4)
			= \bar{Q}_0(\sigma_1, \sigma_2;\sigma_3, \sigma_4)
				+ \int \dd\beta \dd\beta' \dd\gamma \dd\gamma' \, \Big(&
					\bar{G}_{[0]}(\sigma_1, \beta) \bar{G}_{[0]}(\sigma_2, \beta') \bar \Gamma(\beta, \beta'; \gamma, \gamma') \\
					&\times \bar{G}_{[0]}(\gamma, \sigma_3)\bar{G}_{[0]}(\gamma', \sigma_4) \Big).
	\end{aligned}
\end{equation}
We call $\bar F(\sigma_1, \sigma_2;\sigma_3, \sigma_4)$ the second term of the right hand side of \eqref{eq:Qexpression},
\begin{equation}
	\bar F(\sigma_1, \sigma_2;\sigma_3, \sigma_4) = \int \dd\beta \dd\beta' \dd\gamma \dd\gamma'\ \bar{G}_{[0]}(\sigma_1, \beta) \bar{G}_{[0]}(\sigma_2, \beta') \bar \Gamma(\beta, \beta'; \gamma, \gamma') \bar{G}_{[0]}(\gamma, \sigma_3) \bar{G}_{[0]}(\gamma', \sigma_4).
\end{equation}
We re-parametrize $\sigma_1, \sigma_2, \sigma_3, \sigma_4 = f(t_1), f(t_2), f(t_3), f(t_4)$ and $\beta, \beta', \gamma, \gamma' = f(t), f(t'), f(\tau), f(\tau')$ so to get the scaling dimension of $\bar F$.
This leads to
\begin{equation}
	\bar F(\sigma_1, \sigma_2;\sigma_3, \sigma_4) = \frac{\bar F(t_1, t_2; t_3, t_4)}{|f'(t_1)f'(t_2)f'(t_3)f'(t_4)|^{\frac{1}{D + 1}}}.
\end{equation}
This tells us that $\bar F$ indeed scales and the scaling dimension is $\frac{1}{D + 1}$.
This together with the fact that $\bar{Q}_0$ has scaling dimension $1/(D + 1)$ implies that $\bar{Q}$ has scaling dimension $1/(D + 1)$.
Let us compute the scaling of the NLO $2$-point function.
In the large $g$ limit we have that,
\begin{multline}
	\begin{aligned}
	\bar G_{\text{NLO}}(\sigma_1, \sigma_2)
		\approx D\, g^2 \int \dd\gamma \dd\gamma' \dd\beta \dd\beta' \, \Big(&
			\bar G(\sigma_1, \gamma) \bar G(\gamma, \beta)^{D-1} \bar G(\beta', \gamma')^{D-1} \\
			&\times \bar G(\beta, \beta') \bar Q(\gamma, \beta; \gamma', \beta') \bar G(\gamma', \sigma_2)
			\Big)
	\end{aligned}
	\\
	\begin{aligned}
		+ \ \frac{D(D-1)}{2}\, g^2 \int \dd\gamma \dd\gamma' \dd\beta \dd\beta'\, \Big(&
		\bar G(\sigma_1, \gamma) \bar G(\gamma, \gamma')^{D-2} \bar G(\gamma, \beta) \bar G(\gamma', \beta') \\
		&\times \bar G(\beta, \beta')^{D-1} \bar Q(\gamma, \gamma'; \beta, \beta') \bar G(\gamma', \sigma_2)
		\Big)
	\end{aligned}
\end{multline}
which leads after simplifications to
\begin{equation}
	\bar G_{\text{NLO}}(\sigma_1, \sigma_2) = \frac{\bar G_{\text{NLO}}(t_1, t_2)}{|f'(t_1) f'(t_2)|^{\frac{1}{D + 1}}}.
\end{equation} 
This shows that the scaling dimension of $\bar G_{\text{NLO}}$ is $\frac{1}{D + 1}$ as for the leading order term.


\section{Multi-orientable SYK tensor model}
\label{sec:syk-multi-tensor}

The $\group{U}(N) \times \group{O}(N) \times \group{U}(N)$ model has been introduced in the tensor model literature in \cite{Tanasa:2012:MultiorientableGroupField}.
It was called the multi-orientable model.
It has then been stated that it should be related to a complex fermions version of the SYK model in \cite{Klebanov:2017:UncoloredRandomTensors}.
The model is defined as follows.
One consider a pair of complex fermionic tensor fields $\psi, \bar\psi$ of rank $3$.
The partition function of the model writes
\begin{equation}
	Z^{\textrm{m.o.}}_{\lambda, N} = \int \mathcal{D}\psi \mathcal{D}\bar{\psi} \e^{- \int \dd t\, L[\psi]}
\end{equation}
where
\begin{equation}
	L[\psi] = \sum_{n} \bar{\psi}_n \partial_t \psi_n + \frac{\lambda}{N^{3/2}} \sum_{\substack{i, j, k, i', j', k'}} \psi_{ijk}(t) \bar{\psi}_{kj'i'}(t) \psi_{k'ji'}(t) \bar{\psi}_{k'j'i}(t)
\end{equation} 
and we also define
\begin{equation}
	g = \lambda^2.
\end{equation} 
The fields transform under the natural action of $\group{U}(N) \times \group{O}(N) \times \group{U}(N)$ and the action is invariant under this transformation.

\medskip

\begin{figure}
	\centering
	\includegraphics[scale = 0.8]{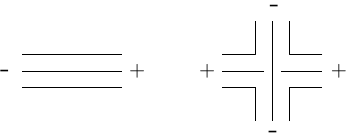}
	\caption{Propagator and vertex of the multi-orientable model.}
	\label{fig:buildingblocksmo}
\end{figure}

The Feynman graphs are constructed out of the building blocks represented on
\cref{fig:buildingblocksmo} with the condition that a $( + )$ half-edge can only connect to a $(-)$ half-edge.
It is also possible to define the notion of jackets for these graphs.
This is indeed a non trivial statement as one can find examples of tensor models for which this is not the case because of the so called \emph{tadface} graphs, see \cite{Tanasa:2016:MultiOrientableRandomTensor, Dartois:2014:1NExpansionMultiorientable} for a discussion of this topics.

Thanks to this notion of jackets the degree can be generalized in this case and it can be shown that the multi-orientable model has a well defined $1/N$ expansion.
So to say we have for the free energy,
\begin{equation}
	F^{\textrm{m.o.}}_{\lambda, N} = \log Z^{\textrm{m.o.}}_{\lambda, N}
		= \sum_{\varpi\ge 0} N^{3-\varpi} F^{\textrm{m.o.}}_{[\varpi]}(\lambda).
\end{equation}
In this case however, $\varpi \in \frac{1}{2}\mathbb{N}_{\ge 0}$, where $\mathbb{N}_{\ge 0}$ is the set of integer larger or equal to zero.
We can again define the two-point function.
Let us start by the free one,
\begin{align}
	&G_f(t_1, t_2) = \frac{1}{N^3} \Big\langle \sum_n T \bar{\psi}_n(t_1) \psi_n(t_2) \Big\rangle_0
		= \sign(t_1 - t_2), \\
	& \Big\langle T \bar{\psi}_{ijk}(t_1) \psi_{i'j'k'}(t_2) \Big\rangle_0 = G_f(t_1, t_2) \, \delta_{ii'} \delta_{jj'} \delta_{kk'},
\end{align} where $n$ here is a multi-index that labels the components of the tensor.
For the exact two-point function we have,
\begin{align}
	& G_e(t_1, t_2) = \frac{1}{N^3} \Big\langle \sum_n T \bar{\psi}_n(t_1)\psi_n(t_2) \Big\rangle, \\ 
	& \Big\langle T \bar{\psi}_{ijk}(t_1) \psi_{i'j'k'}(t_2) \Big\rangle = G_e(t_1, t_2) \, \delta_{ii'} \delta_{jj'} \delta_{kk'}.
\end{align}
Consequently we have,
\begin{equation}
	G_e(t_1, t_2) = \sum_{\varpi \in \frac{1}{2} \mathbb{N}} N^{-\varpi} G_{[\varpi]}(t_1, t_2).
\end{equation}

\subsection{The Leading Order}

As was shown in \cite{Dartois:2014:1NExpansionMultiorientable}, the leading order in $N$ is once again dominated by melonic graphs.
As a consequence we can write the equation satisfied by the two-point function at leading order,
\begin{equation}\label{eq:exactLOmo}
	G_{[0]}(t_1, t_2) = G_f(t_1, t_2) + g \int \dd t \dd t' \, G_f(t_1, t)G_{[0]}(t, t')^3G_{[0]}(t', t_2).
\end{equation}
Using now known manipulations we have in the infrared/large coupling limit the approximated equation
\begin{equation}
	g \int \dd t \, \bar{G}_{[0]}(t_1, t)^3 \bar{G}_{[0]}(t, t_2) = - \delta(t_1 - t_2).
\end{equation} 
This equation has a known solution
\begin{equation}
	\bar{G}_{[0]}(t_1, t_2) = \left(\frac{\tan(\pi/4)}{4\pi g}\right)^{1/4} \frac{\sign(t_1 - t_2)}{|t_1 - t_2|^{1/2}}.
\end{equation} 
Moreover, the large coupling equation has the same re-parametrization symmetry.
If $\bar{G}_{[0]}(t_1, t_2)$ is a solution, then, $\bar{G}_{[0]}(\sigma_1, \sigma_2)$, where $\sigma_{1, 2} = f(t_{1, 2})$, is a solution as well, provided that $\bar{G}_{[0]}(t_1, t_2) = |\partial_{t_1} f(t_1) \partial_{t_2} f(t_2)|^{\frac{1}{4}} \bar{G}_{[0]}(\sigma_1, \sigma_2)$.

\subsection{The Next-to-Leading Order}

The next-to-leading order of the two-point function of the multi-orientable model has been studied in \cite{Raasakka:2015:NexttoleadingOrderLarge}.
As the combinatorics is unchanged by the fact that we consider fermionic fields on one dimensional space we can easily infer the next-to-leading order in this case.
The degree at next-to-leading order is
\begin{equation}
	\varpi_{\text{NLO}} = \frac{1}{2}.
\end{equation}
The self-energy $\Sigma_{\text{NLO}}(t_1, t_2) := \Sigma_{[1/2]}(t_1, t_2)$ at next-to-leading order writes graphically,
\begin{equation}
	\Sigma_{\text{NLO}}(t_1, t_2) = \ \raisebox{-3.0ex}{\includegraphics[scale = 0.6]{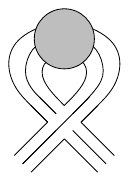}} \ + \ \raisebox{-9.0ex}{\includegraphics[scale = 0.85]{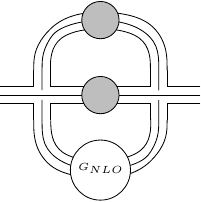}}
\end{equation}
where the gray disks represent insertion of the leading order two-point function on the edges.
This translates into the formal equation
\begin{equation}\label{eq:self-mo-equation}
	\Sigma_{\text{NLO}}(t_1, t_2) = \lambda \, \delta(t_1 - t_2) \, G_{[0]}(t_1, t_2) + 3 g \, G_{[0]}(t_1, t_2)^2G_{[1/2]}(t_1, t_2).
\end{equation}
We also have,
\begin{equation}\label{eq:selftoGmo}
	G_{\text{NLO}}(t_1, t_2) := G_{[1/2]}(t_1, t_2) = \int \dd t \dd t' \, G_{[0]}(t_1, t)\Sigma_{\text{NLO}}(t, t')G_{[0]}(t', t_2).
\end{equation}

We now use the analogue of the operator\footnotemark{} $K(t_1, t_3; t_3, t_4)$ introduced in earlier sections.
\footnotetext{%
	Notice however the difference in sign.
}%
It is here defined as
\begin{equation}
	K(t_1, t_3;t_3, t_4) = 3 g \, G_{[0]}(t_1, t_3) G_{[0]}(t_2, t_4) G_{[0]}(t_3, t_4)^2.
\end{equation} 
We introduce $G_{[0]}(t_1, t_2) = G_{[0]}(t_2-t_1) = G_{[0]}(\tau)$, with $\tau = t_2 - t_1$.
Then, using \eqref{eq:selftoGmo} and \eqref{eq:self-mo-equation}, we can write at least formally,
\begin{align}
	G_{\text{NLO}}(t_1, t_2) &= \lambda \int \dd t \, G_{[0]}(t_1, t) G_{[0]}(0) G_{[0]}(t, t_2) + \int \dd t \dd t' \, K(t_1, t_2; t, t') G_{\text{NLO}}(t, t')\\
		&= \lambda \int \dd t \, G_{[0]}(t_1, t)G_{[0]}(0) G_{[0]}(t, t_2) + \left[K\ast G_{\text{NLO}} \right](t_1, t_2),
\end{align}
and thus
\begin{align}
	\left[(\delta^{\otimes 2} - K) \ast G_{\text{NLO}} \right](t_1, t_2)& := \int \dd t \dd t' \, \big(\delta(t_1 - t) \otimes \delta(t_2-t') - K(t_1, t_2;t, t')\big) G_{\text{NLO}}(t, t') \\
		&= \lambda \int \dd t G_{[0]}(t_1, t) G_{[0]}(0) G_{[0]}(t, t_2).
\end{align}
However, since fermionic two-point functions are anti-symmetric in the time variables, we have\footnotemark{} $G_{[0]}(0) = 0$, which then implies
\footnotetext{%
	One can also convince oneself by going into Fourier space and defining an appropriate cut-off for the regularization of the integral.
}%
\begin{equation}
	\left[(\delta^{\otimes 2} - K)\ast G_{\text{NLO}}\right](t_1, t_2) = 0, \qquad \forall t_1, t_2.
\end{equation}
This implies that $G_{\text{NLO}}$ must lie in the kernel of $(\delta^{\otimes 2} - K)$.
This happens as long as $G_{\text{NLO}} = 0$ or $G_{\text{NLO}}$ is an eigenvector of $K$ with eigenvalue $1$.
Since there are such eigenvectors $\bar{G}_{\text{NLO}}$ can be an arbitrary linear combination of them if it does not vanish and, without additional data on the behaviour of $\bar{G}_{\text{NLO}}$ it is not possible to conclude about its conformality.

\section{Discussion}
\label{sec:discussion}

We have found that the NLO in the large $N$ expansion does not modify the dependence of the $2$-point function in the coupling and time in the infrared regime.
For this reason the $2$-point function is still conformally invariant and the IR dimension of the fermions does not receive any correction at this order.
Nonetheless higher-order correlation functions may deviate from the CFT behaviour and this provides an incentive to study their behaviour.
In any case one can consider the NLO as being a CFT in any context where the corrections to these higher-order functions can be neglected.

This fact may reveal itself to be important in the construction of the bulk dual using the AdS/CFT dictionary~\cite{Gross:2017:BulkDualSYK}: absence of $1/N$ corrections in the CFT translates into absence of quantum corrections in the bulk dual.
For example if the scaling dimensions of the single traces operators discussed in~\cite{Maldacena:2016:CommentsSachdevYeKitaevModel, Gross:2017:BulkDualSYK} are identical at NLO this would translate by the fact that the corresponding bulk field masses do not receive correction at one loop.
Hence our result gives a strong indication that the first quantum correction may be absent and this point calls for a deeper study.

A natural extension of this work would be to determine how the spontaneous breaking of the conformal symmetry appears in the NLO $4$-point function and what are the effects of incorporating the NLO correction of the coupling constant.
Another point of interest is to push the study even further and see if the NNLO continues to preserve the conformal invariance.
The method described in this paper can be generalized to study the NLO in other models, such as in the supersymmetric case~\cite{Fu:2017:SupersymmetricSYKModels, Peng:2016:SupersymmetricSYKlikeTensor}.
Finally it would be useful to settle the question of the conformal invariance of the $2$-point function in the multi-orientable tensor model.

\section*{Acknowledgments}

We are grateful to Costas Bachas and Ashoke Sen for useful discussions.
The work of S.M., made within the \textsc{Labex Ilp} (reference \textsc{Anr–10–Labx–63}), is supported by French state funds managed by the \emph{Agence nationale de la recherche}, as part of the program \emph{Investissements d'avenir} under the reference \textsc{Anr–11–Idex–0004–02}.
S.M. is supported by Cefipra under project 5204-4.

\appendix

\section{Composite field effective action}
\label{sec:fluctuations}

The goal of this section is to introduce composite fields for colored tensor models.
The original motivation was to find an effective action in terms of composite fields to study the IR regime of colored tensor models.
However, the naive approach proposed here does not work because it is not suited for an IR approximation (see the discussion at the end of \Cref{sec:fluctuations:eff-action}).
Nonetheless, we find useful to provide the details as an illustration.\footnotemark{}
\footnotetext{%
	Which also motivates the need for the more advanced machinery developed in~\cite{Benedetti:2018:2PIEffectiveAction} which appeared after our paper.
}%

We will focus on real tensors for simplicity but the generalization to complex tensors is straightforward.

Recall the action for $D + 1$ real fermionic tensor fields (\cref{sec:syk-colored-tensor})
\begin{equation}
	\label{eq:action:syk-tensor}
	S[\psi^c]
		= \int \dd t \left( \frac{1}{2} \sum_c \psi^c \pd_t \psi^c + \I^{\frac{D+1}{2}} \Lambda\, \prod_c \psi^c \right),
\end{equation}
where the contraction over the tensor indices is implicit and $c = 0, \ldots, D$ and $\Lambda$ is defined by
\begin{equation}
	\label{eq:relation-g-lambda}
	\Lambda
		= \frac{\lambda}{N^{\frac{D (D - 1)}{4}}}
\end{equation}
The associated partition function is
\begin{equation}
	\label{eq:path-integral}
	Z
		= \int \prod_c \mathcal{D}\psi^c\, \e^{- S[\psi^c]}.
\end{equation}

\subsection{Effective action}
\label{sec:fluctuations:eff-action}

In order to introduce a composite field\footnotemark{}
\footnotetext{%
	The composite fields are distinguished from the correlation functions by the absence of any lower index.
}%
\begin{equation}
	\label{eq:corr:2point-exact}
	G_{n_c, n'_c}^c(t, t')
		= - \psi_{n_c}^c(t) \psi_{n'_c}^c(t')
\end{equation}
where $n_c$ and $n'_c$ are tensor multi-indices, corresponding to the $2$-point function
\begin{equation}
	G_e^c(t, t')
		= - \frac{1}{N^D} \mean{\sum_{n_c} \psi_{n_c}^c(t) \psi_{n_c}^c(t')},
\end{equation}
one first needs to obtain an action with bilinear terms in each color.
In the rest of this section, the tensor indices will be implicit.

This can be achieved by integrating out one of the color, say $\psi^0$ which is straightforward since the action is quadratic in this field, the product
\begin{equation}
	\Psi
		= \I^{\frac{D+1}{2}} \Lambda \prod_{i = 1}^D \psi^i
\end{equation}
acting as a source for $\psi^0$, where $i = 1, \ldots, D$.
Using standard techniques, the effective action obtained after integrating out $\psi^0$ is
\begin{equation}
	\label{eq:action:syk-tensor-0-integrated}
	S_{\text{eff}}[\psi^i]
		= \frac{1}{2} \sum_i \int \dd t\, \psi^i \pd_t \psi^i
			+ \I^{D^2+1}\, \frac{\Lambda^2}{2} \int \dd t \dd t'\, S(t, t') \prod_i \psi^i(t) \psi^i(t').
\end{equation}
after rearranging the fermions (the signs have been traded for $\I$), where $S(t, t')$ is the Green function for $\pd_t$.

The next step consists in introducing the bilocal tensor fields $G^i(t, t')$ \eqref{eq:corr:2point-exact} and to use auxiliary fields $\Sigma^i(t, t')$ such that
\begin{subequations}
\begin{align}
	1
		&= \int \prod_i \mathcal{D} G^i\, \delta\big(G^i(t, t') + \psi^i(t) \psi^i(t') \big)
		\\
		&= \int \prod_i \mathcal{D} G^i \mathcal{D} \Sigma^i\, \e^{- S_{\text{aux}}[\psi^i, G^i, \Sigma^i]}
\end{align}
\end{subequations}
where
\begin{equation}
	S_{\text{aux}}[\psi^i, G^i, \Sigma^i]
		= - \frac{1}{2} \sum_i \int \dd t \dd t'\,
			\Sigma^i(t, t') \big(G^i(t, t') + \psi^i(t) \psi^i(t') \big).
\end{equation}
The functional integral \eqref{eq:path-integral} becomes
\begin{equation}
	Z
		= \int \prod_i \mathcal{D} \psi^i \mathcal{D} G^i \mathcal{D} \Sigma^i\,
			\e^{- \tilde S_{\text{eff}}[\psi^i, G^i] - S_{\text{aux}}[G^i, \Sigma^i]}.
\end{equation}
where
\begin{equation}
	\tilde S_{\text{eff}}[\psi^i, G^i]
		= \frac{1}{2} \sum_i \int \dd t\, \psi^i \pd_t \psi^i
			+ \I^{(D + 1)^2}\, \frac{\Lambda^2}{2} \int \dd t \dd t'\, S(t, t') \prod_i G^i(t, t').
\end{equation}
Performing the quadratic integration over $\psi^i$ yields the effective action for $G^i$ and $\Sigma^i$
\begin{equation}
	\label{eq:action:G-Sig}
	\begin{aligned}
		W[G^i, \Sigma^i]
			= &- \frac{1}{2} \sum_i \tr \ln(\pd_t - \Sigma^i)
				- \frac{1}{2} \sum_i \int \dd t \dd t'\, \Sigma^i(t, t') G^i(t, t')
				\\
				&+ \I^{(D + 1)^2}\, \frac{\Lambda^2}{2} \int \dd t \dd t'\, S(t, t') \prod_i G^i(t, t').
	\end{aligned}
\end{equation}

The equations of motion are
\begin{subequations}
\label{eq:saddle-point}
\begin{align}
	\label{eq:saddle-point-G}
	\frac{\delta W}{\delta G^i}
			&= 0
		\quad \Longrightarrow \quad
		\Sigma^i(t, t')
			= \I^{(D + 1)^2} \Lambda^2\, S(t, t') \prod_{j \neq i} G^j(t, t'),
	\\
	\label{eq:saddle-point-Sig}
	\frac{\delta W}{\delta \Sigma^i}
			&= 0
		\quad \Longrightarrow \quad
		\Big(\delta(t - t') \mathbbm{1}^{\otimes D} \pd_t - \Sigma^i(t, t') \Big)^{-1} - G^i(t, t')
			= 0
\end{align}
\end{subequations}
where $\mathbbm{1}^{\otimes D}$ is the tensor identity.
The last equation can be rewritten as
\begin{subequations}
\label{eq:saddle-point-Sig-integral}
\begin{align}
	\delta(t - t'')\, \mathbbm{1}^{\otimes D}
		&= \pd_t G^i(t, t'') + \int \dd t'\, \Sigma^i(t, t') G^i(t', t''),
		\\
		&= \pd_t G^i(t, t'') + \I^{(D + 1)^2} \Lambda^2 \int \dd t'\, S(t, t') G^i(t', t'') \prod_{j \neq i} G^j(t, t')
\end{align}
\end{subequations}
where the last equality follows from inserting \eqref{eq:saddle-point-G}.

The computations in this section are exact, which means that the effective action \eqref{eq:action:G-Sig} is exact and leads to the correct the Schwinger--Dyson equations.\footnotemark{}
\footnotetext{%
	This paragraph is an answer to~\cite[p.~2]{Benedetti:2018:2PIEffectiveAction} which states that the action derived here is not correct because it gives “wrong Schwinger--Dyson equations”.
	Note also that the action from~\cite{Benedetti:2018:2PIEffectiveAction} (see also~\cite{Choudhury:2018:NotesMelonicONq1}) is obtained by making approximations, implying that it is not exact.
}%
The drawback of our action (and of the corresponding Schwinger--Dyson equations) over the one in~\cite{Benedetti:2018:2PIEffectiveAction} is that it is not suited for studying the IR regime.
Indeed, they implicitly include all modes from the $\psi^0$ field (and thus the UV fluctuations) because it has been exactly integrated out.
By contrast in the usual SYK, we integrate out the random coupling constants which are non-dynamical (they are pure IR): this explains why there is no difficulty in taking the IR limit afterwards, because there is no hidden UV contribution as we get when integrating out a dynamical field.
In principle, we could correct our action through some kind of renormalization group approach in order to integrate only the IR modes of the field $\psi^0$.

\subsection{Fluctuations}

The solutions to the equations of motion are denoted by $(G_{[0]}, \Sigma_{[0]})$ and they are identical for all colors since the equations are symmetric under exchange of colors
\begin{equation}
	\mean{G^i}
		= G_{[0]}\, \mathbbm{1}^{\otimes D}, \qquad
	\mean{\Sigma^i}
		= \Sigma_{[0]}\, \mathbbm{1}^{\otimes D},
\end{equation}
where $G_{[0]}$ and $\Sigma_{[0]}$ are genuine bilocal fields (not tensors).
Powers of $\mean{G^i}$ (or $\mean{\Sigma^i}$) will be accompanied by factors of $N$ due to the contraction of the identities
\begin{equation}
	(\mean{G^i})^k
		= N^{k + \binom{k}{2}} G_{[0]}^k
		= N^{\frac{k (k+1)}{2}} G_{[0]}^k.
\end{equation}

The saddle point equations \eqref{eq:saddle-point} become
\begin{subequations}
\begin{align}
	\label{eq:saddle-point-G0}
	\Sigma_{[0]}(t, t')
		= \I^{(D + 1)^2} \lambda^2\, S(t, t') G_{[0]}(t, t')^{D-1}, \\
	\label{eq:saddle-point-Sig0}
	\big( \delta(t - t') \pd_t + \Sigma_{[0]}(t, t') \big)^{-1} - G_{[0]}(t, t')
		= 0.
\end{align}
\end{subequations}
where the relation \eqref{eq:relation-g-lambda} between $\Lambda$ and $\lambda$ has been used.

Then one can consider fluctuations $(g^i, \sigma^i)$ around these solutions
\begin{equation}
	G^i
		= G_{[0]}\, \mathbbm{1}^{\otimes D} + g^i, \qquad
	\Sigma^i
		= \Sigma_{[0]}\, \mathbbm{1}^{\otimes D} + \sigma^i.
\end{equation}
Plugging these expressions into \eqref{eq:action:G-Sig} yield
\begin{equation}
	\label{eq:action:g-sig}
	\begin{aligned}
		W[g^i, \sigma^i]
			=\ &\frac{1}{4} \sum_i \int \dd t_1 \cdots \dd t_4\, \sigma^i(t_1, t_2) k(t_1, \ldots, t_4) \sigma^i(t_3, t_4)
				- \frac{1}{2} \sum_i \int \dd t \dd t'\, \sigma^i g^i
				\\
				&+ \I^{(D + 1)^2}\, \frac{\lambda^2}{4 N^{D-1}} \int \dd t \dd t'\, S\, G_{[0]}^{D-2} \sum_{i, j} g^i g^j
				+ \frac{1}{2} \sum_i \sum_{n \ge 3} \frac{1}{n}\, \tr (G_{[0]} \sigma^i)^n
				\\
				&+ \frac{\I^{(D + 1)^2}}{N^{\frac{D (D-1)}{2}}}\, \sum_{n=3}^D \frac{\lambda^2}{2 n!}\, N^{\frac{(D-n) (D-n+1)}{2}} \int \dd t \dd t'\, S\, G_{[0]}^{D-n} \sum_{i_1, \ldots, i_n} g^{i_1} \cdots g^{i_n}
	\end{aligned}
\end{equation}
where the dependence in the time $(t, t')$ has been omitted and the kernel $k$ is
\begin{equation}
	\label{eq:def-K-op}
	k(t_1, \ldots, t_4)
		= G_{[0]}(t_1, t_3) G_{[0]}(t_2, t_4).
\end{equation}
Rescaling the fluctuations such that
\begin{equation}
	G^i
		= G_{[0]}\, \mathbbm{1}^{\otimes D} + \abs{G_{[0]}}^{\frac{1-D}{2}} g^i, \qquad
	\Sigma^i
		= \Sigma_{[0]}\, \mathbbm{1}^{\otimes D} + \abs{G_{[0]}}^{\frac{D-1}{2}} \sigma^i
\end{equation}
and absorbing the factors inside the kernel gives the symmetric kernel~\cite{Maldacena:2016:CommentsSachdevYeKitaevModel}
\begin{equation}
	K_{\text{sym}}(t_1, \ldots, t_4)
		= - \lambda^2\, \abs{G_{[0]}(t_1, t_2)}^{\frac{D-1}{2}} k(t_1, \ldots, t_4) \abs{G_{[0]}(t_3, t_4)}^{\frac{D-1}{2}}
\end{equation}
which is conjugated to the kernel \eqref{eq:operator-K}.

Truncating the action to the quadratic order, one can obtain an effective action for the $g^i$ only by integrating out $\sigma^i$
\begin{equation}
	W_{\text{eff}}[g^i]
		= - \frac{\lambda^2}{4} \sum_{i,j}
			\int \dd t_1 \cdots \dd t_4\, g^i(t_1, t_2) \mc K^{ij}(t_1, \ldots, t_4) g^j(t_3, t_4)
\end{equation}
where
\begin{equation}
	\mc K^{ij}(t_1, \ldots, t_4)
		= K_{\text{sym}}^{-1}(t_1, \ldots, t_4) \delta^{ij}
			- \frac{\I^{(D + 1)^2}}{N^{D-1}}\, S(t_1, t_2)
				\delta(t_1 - t_3) \delta(t_2 - t_4).
\end{equation}

The $4$-point function for the fermions correspond to the $2$-point function of the fluctuations $\mean{g^i g^j}$.
At leading order it can be computed using the above quadratic action and one can see that it is equivalent to the computation done in \cref{sec:syk-colored-tensor}.
The additional propagator in the action is a consequence of integrating out one of the colors and it is present to connect vertices, very similar to the way one adds an extra line after averaging over disorder in the standard SYK model~\cite{Maldacena:2016:CommentsSachdevYeKitaevModel}.
However this time the extra line represents a dynamical fields.

\printbibliography[heading=bibintoc]

\end{document}